\documentclass[8.5pt,twoside,twocolumn]{article}
\usepackage[super,sort&compress,comma]{natbib} 
\usepackage{times,mathptmx}

\usepackage{sectsty}
\usepackage{balance} 

\usepackage[pdftex]{graphicx} 
\usepackage{lastpage}
\usepackage[format=plain,justification=raggedright,singlelinecheck=false,font=small,labelfont=bf,labelsep=space]{caption} 
\usepackage{fancyhdr}
\usepackage{color}

\begin{document}

\thispagestyle{plain}
\fancypagestyle{plain}{
\fancyhead[L]{\includegraphics[height=8pt]{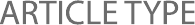}}
\fancyhead[C]{\hspace{-1cm}\includegraphics[height=20pt]{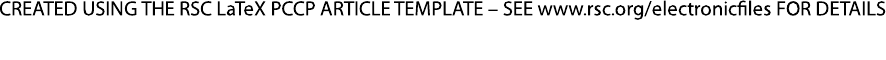}}
\fancyhead[R]{\includegraphics[height=10pt]{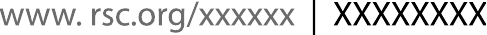}\vspace{-0.2cm}}
\renewcommand{\headrulewidth}{1pt}}
\renewcommand{\thefootnote}{\fnsymbol{footnote}}
\renewcommand\footnoterule{\vspace*{1pt}%
\hrule width 3.4in height 0.4pt \vspace*{5pt}} 
\setcounter{secnumdepth}{5}

\makeatletter 
\def\subsubsection{\@startsection{subsubsection}{3}{10pt}{-1.25ex plus -1ex minus -.1ex}{0ex plus 0ex}{\normalsize\bf}} 
\def\paragraph{\@startsection{paragraph}{4}{10pt}{-1.25ex plus -1ex minus -.1ex}{0ex plus 0ex}{\normalsize\textit}} 
\renewcommand\@biblabel[1]{#1}            
\renewcommand\@makefntext[1]%
{\noindent\makebox[0pt][r]{\@thefnmark\,}#1}
\makeatother 
\renewcommand{\figurename}{\small{Fig.}~}
\sectionfont{\large}
\subsectionfont{\normalsize} 

\fancyfoot{}
\fancyfoot[LO,RE]{\vspace{-7pt}\includegraphics[height=9pt]{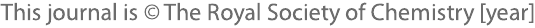}}
\fancyfoot[CO]{\vspace{-7.2pt}\hspace{12.2cm}\includegraphics{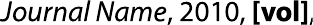}}
\fancyfoot[CE]{\vspace{-7.5pt}\hspace{-13.5cm}\includegraphics{RF}}
\fancyfoot[RO]{\footnotesize{\sffamily{1--\pageref{LastPage} ~\textbar  \hspace{2pt}\thepage}}}
\fancyfoot[LE]{\footnotesize{\sffamily{\thepage~\textbar\hspace{3.45cm} 1--\pageref{LastPage}}}}
\fancyhead{}
\renewcommand{\headrulewidth}{1pt} 
\renewcommand{\footrulewidth}{1pt}
\setlength{\arrayrulewidth}{1pt}
\setlength{\columnsep}{6.5mm}
\setlength\bibsep{1pt}

\twocolumn[
  \begin{@twocolumnfalse}
\noindent\LARGE{\textbf{Experimental study of forces between
quasi-two-dimensional emulsion droplets near jamming}}
\vspace{0.6cm}

\noindent\large{\textbf{Kenneth W. Desmond$^a$, Pearl J. Young,
Dandan Chen$^{b}$, Eric R. Weeks$^\ast$}}\vspace{0.5cm}



\noindent \normalsize{We experimentally study the jamming of quasi-two-dimensional emulsions.  Our experiments consist of oil-in-water emulsion droplets confined between two parallel plates. From the droplet outlines, we can determine the forces between every droplet pair to within 8\% over a wide range of area fractions $\phi$.  We study three bidisperse samples that jam at area fractions $\phi_c \approx 0.86$. Our data show that for $\phi > \phi_c$,  the contact numbers and pressure have power-law dependence on $\phi-\phi_c$ in agreement with the critical scaling found in numerical simulations.  Furthermore, we see a link between the interparticle force law and
the exponent for the pressure scaling, supporting prior computational
observations. We also observe linear-like force chains (chains of large inter-droplet forces) that extend over 10 particle lengths, and examine the origin of their linearity. We find that the relative orientation of  large force segments are random and that the tendency for force chains to be linear is not due to correlations in the direction of neighboring large forces, but instead occurs because the directions  are biased towards being linear to balance the forces on each droplet.
}
\vspace{0.5cm}
 \end{@twocolumnfalse}
  ]



\footnotetext{\textit{Department of Physics, Emory University,
Atlanta, GA 30322, USA.}}
\footnotetext{\textit{$^{a}$~Current address:  Department of
Mechanical Engineering, University of California, Santa Barbara,
CA 93106, USA.}}
\footnotetext{\textit{$^{b}$~Current address:  Soochow
University, Suzhou, Jiangsu, China}}



\section{Introduction}

A liquid to amorphous-solid transition, also known as a jamming
transition, occurs in a wide variety of soft materials such as
colloids, emulsions, foams, and sand.  In general the jamming
transition is due to an increase in the particle concentration
$\phi$;  the particles become sufficiently crowded so that
microscopic rearrangements are unable to occur when external
stresses are applied~\cite{Trappe2001, Siemens2010, Hecke2010}. At
a critical $\phi_c$ the system jams into a rigid structure,
and many of the material properties are known \cite{Liu2010,
Siemens2010} to scale with a power-law dependence on $(\phi -
\phi_c)$.  While these soft materials have obvious differences,
it has been postulated that there are universal features of the
jamming transition that all these materials share in common such
as critical scaling and the emergence of force chains.

In all systems above the jamming point, particles press into one
another and deform. As the density increases, new contacts form
and particles deform more, increasing the pressure.  Interesting,
both the average number of contacts $z$ and the pressure $P$
show critical-like scaling relative to the jamming point. In
experiments and simulations, both 2D and 3D, the average number
of contacts scales as $z - z_c = A(\phi - \phi_c)^{\beta_z}$,
where $z_c$ and $A$ depend on the dimension and $\beta_z = 1/2$
regardless of dimension\cite{Durian1995, Durian1997, OHernPRL2002,
OHern2003, Ellenbroek2006, Majmudar07, Katgert2010_2}. Simulations
found $P \sim (\phi - \phi_c)^{\beta_P}$, where $\beta_P$ depends
on the details on the interparticle force law~\cite{Durian1995,
Durian1997, OHernPRL2002, OHern2003}. If this pressure scaling and
connection between between $\beta_P$ and the interparticle force
law extends to experiments, then this would demonstrate a direct
link between the interaction of the constituent particles and the
bulk properties of the sample, as the bulk modulus can be found
from $P(\phi)$.

Another observed feature of jammed systems is the spatial
heterogeneity of the particle-particle contact forces. In
experiments and simulations, both 2D and 3D, the shape of the
probability distribution of forces is broad with an exponential
like tail~\cite{Liu1995, Cates1999, Majmudar07, Majmudar05,
Zhou2006, Howell1999, Howell1999_2, Sun2011, OHernPRL2002,
OHern2003, Brujic2003_2, Brujic2003, JaegerRevModernPhys1996,
Katgert2010_2}. The largest forces
tend to form chain structures that bear the majority of the
load~\cite{Liu1995, Cates1999, Majmudar05, Zhou2006, Howell1999,
Howell1999_2, Sun2011}. These force chains are responsible for
providing rigidity of jammed materials to external stresses and
are related to many other bulk properties~\cite{Tordesillas2007,
Liu1995, Cates1999, Cates1998}. In prior experiments on 3D
emulsions, the structure of the force chains was studied
directly, where force chains extended over 10 particle
diameters with an persistence length of 3 - 4 particle
diameters~\cite{Brujic2003_2, Brujic2003, Zhou2006}.

There have been theoretical attempts to understand
force chains, such as the q-model of Coppersmith {\it
et al.} \cite{Coppersmith1996}, directed-force chain
networks of Socolar's group \cite{Otto2003}, and simulations
\cite{Radjai1996,Thornton1997,OHern2001}.  Others took an ensemble
approach to describe force chains, with different choices for
ensembles~\cite{Brujic2003, Snoeijer2004, Henkes2007, Tighe2008,
Zhou2009, Chakraborty2010, Claudin1998, Edwards1996}.  While some
of these models successfully predict certain properties of the
force network, they can not explain the physical origins of force
chains. To explain the structure of the force chains observed in 3D
emulsion studies, Bruji\'{c} \textit{et al.} ~\cite{Brujic2003_2,
Brujic2003} and Zhou \textit{et al.}~\cite{Zhou2006, Zhou2009}
proposed an accurate model that provides a physical description for
the origin of force chains.  This model has two simple assumptions:
first, the forces on a droplet must balance, and second, forces
between neighboring droplets are uncorrelated.  This model has
not been applied to 2D systems.

In this paper, we introduce a new experimental system to study the
universal nature of the jamming transition. Our system consist of
quasi-2D soft deformable droplets with no static friction forces.
In the appendix, we describe our method to determine the forces
between droplets in contact to within 8\%, significantly better
than prior studies of foams \cite{Katgert2010_2} and comparable
to photoelastic disks \cite{Majmudar05}. Using our experimental
model system, we find power-law scaling for the coordination
number and pressure (Sec.~\ref{critscaling}), we observe a
relationship between the interparticle force law and $\beta_P$
(Sec.~\ref{critscaling}), and see a distribution of contact forces
similar to prior work (Sec.~\ref{forcedistribution}).  Further,
we confirm the assumptions of the Bruji\'{c}-Zhou model apply
to our data and that the model well-describes our 2D data
(Sec.~\ref{forcechains}).  This work provides an in depth study
comparing data from our experimental model system to other numerical
simulations, theory, and experimental systems, thus furthering
our understanding of the jamming transition and supporting the
applicability of ideas of jamming to a new system.

\section{Experimental method}
\label{method}

\begin{figure}
\begin{center}
\includegraphics[width=3.4in]{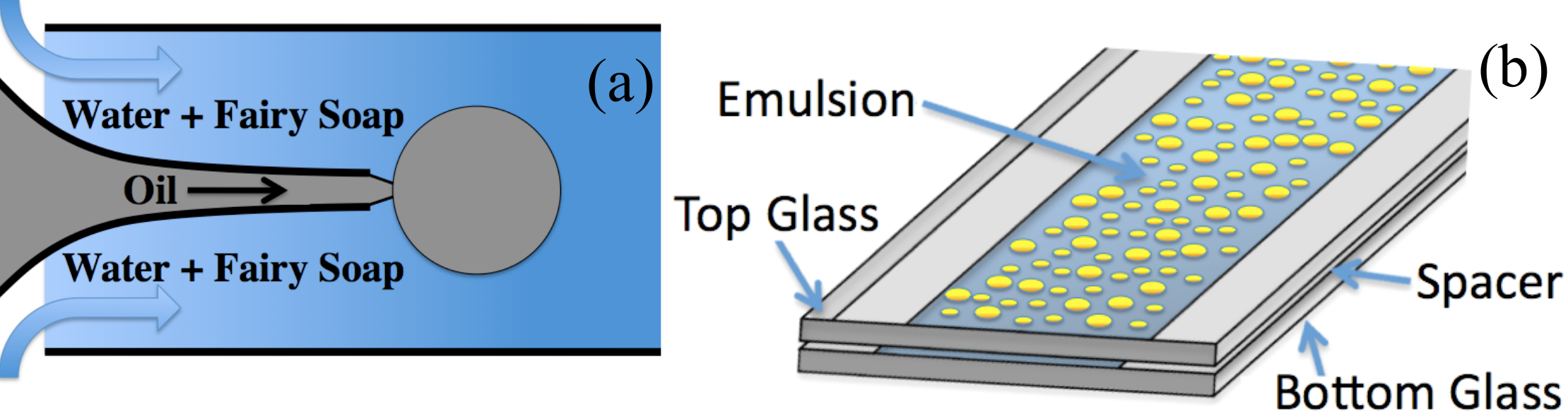}
\caption{(Color online) (a) A schematic of our co-flow apparatus.
Oil is pumped at a constant rate through a micropipet
centered within a capillary tube of larger diameter. Around the
inner micropipet, a 5 g/mL water Fairy soap mixture is pumped
through the capillary tube, and as oil leaves the micropipet it
forms spherical droplets that repeatedly break off with the same
diameter. (b) A schematic of our sample chamber where emulsion
droplets are confined to a 2D plane by two microscope slides
separated by either a $\sim 100$ $\mu$m spacer (transparency film)
or $\sim 180 $ $\mu$m spacer (glass coverslip).}
\label{fig:Sample}
\end{center}
\end{figure}

We produce emulsions using a standard co-flow micro-fluidic
technique~\cite{Shan04}, see Fig.~\ref{fig:Sample}(a). The inner
tube diameter is $\sim 35$~$\mu$m and the outer tube diameter is
$\sim 500$~$\mu$m. 
The continuous phase is a mixture of water and the commercial soap
``Fairy'', and flows through the outer tube at a rate of $\sim
1$ mL/min.  The droplets are mineral oil, which flows through the
inner tube at a rate of $\sim 0.5$ $\mu$L/hr.
Slight variations of these parameters let us produce monodisperse
droplets with radii in the range of 80-170~$\mu$m; any given batch
of droplets has a polydispersity of less than 4\%.  Mixing together
two monodisperse batches lets us produce bidisperse samples with
whatever size ratio and number ratio desired.

Our sample chamber is designed to create a system of quasi-2D
emulsion droplets, analogous to 2D granular systems of
photoelastic disks \cite{Majmudar05} but without static friction.
The chamber consists of two microscope glass slides of dimensions
25 mm $\times$ 75 mm (Corning) separated by a $\sim 100$
$\mu$m spacer (transparency film) or a $\sim 180$ $\mu$m spacer
(Corning No. 1 glass coverslip) glued along the two longer edges;
see Fig.~\ref{fig:Sample}(b).  The sample chamber thickness is
tuned so that the droplets are deformed into pancake shapes,
with aspect ratio (diameter/height) ranging from 1.6 to 3.0;
see Fig.~\ref{fig:finding_R2}.

After the sample chambers are filled, they are placed on a
microscope for imaging with either a $1.6\times$ or $5\times$
objective lens.  The droplets are allowed to equilibrate their
positions; we only consider static samples.  Our camera takes
$2,200 \times 1,800$ pixel$^2$ images.  We overlap images
from different areas to construct a single large field of view
image on the order of $10,000 \times 50,000$ pixel$^2$ containing
between 1,000 to 5,000 droplets depending on the droplet sizes.
We image every droplet (wall-to-wall) and only analyze
droplets more than $\sim 4$ diameters away from the nearest wall
to avoid wall effects \cite{Desmond2009}.

\section{Empirical Force Law}\label{sec:EmpiricalForceLaw}

\begin{figure}
\includegraphics[width=3.2in]{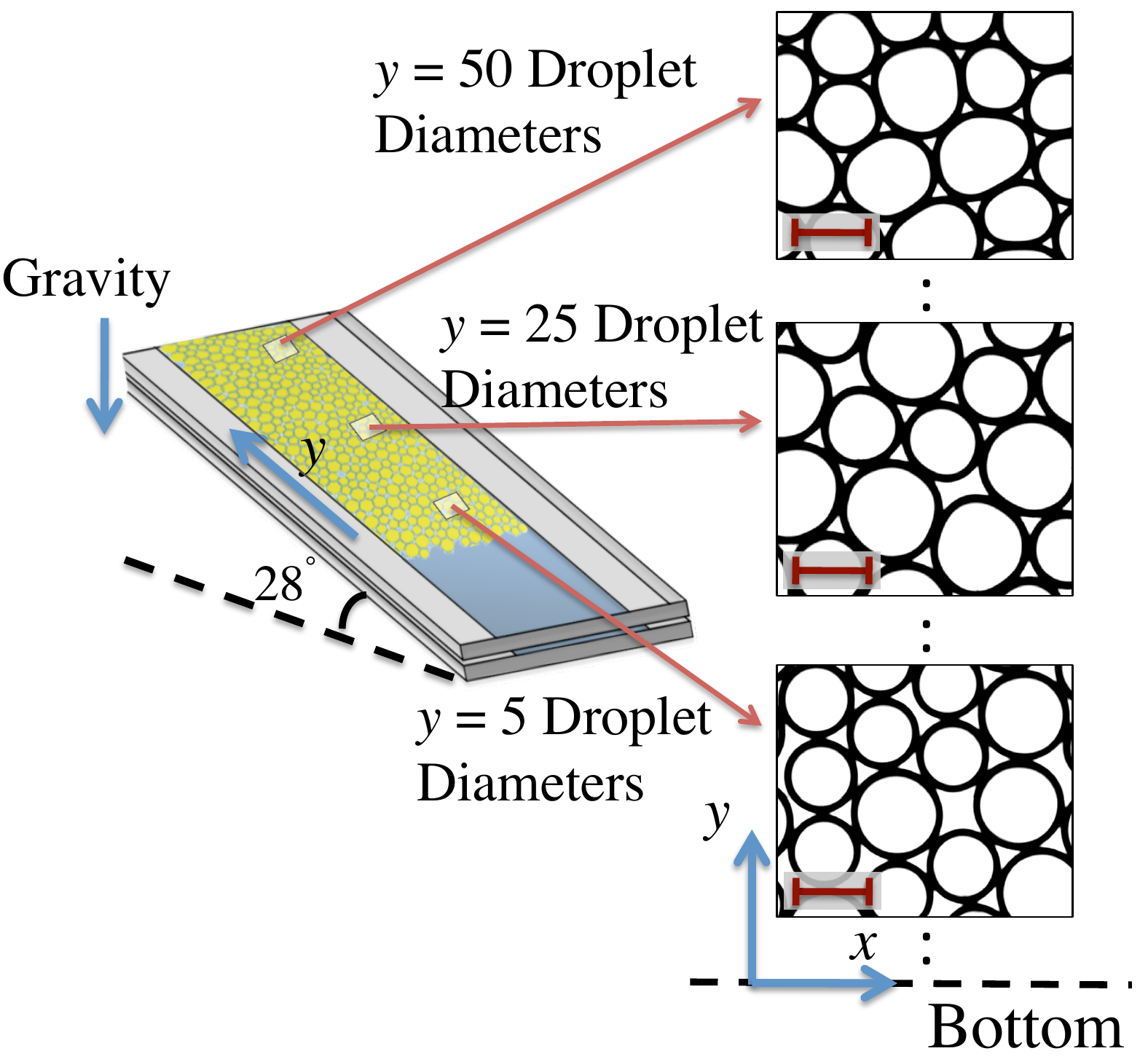}
\caption{(Color online).  Illustration of our experiment.
Oil droplets rise to the top of the sample
chamber due to buoyancy.  At the bottom of the droplet ``pile,''
droplets barely touch and are not deformed.  At the top, droplets
are compressed due to the buoyant weight of the droplets below
them.  This lets us study the sample from the jamming area
fraction on up, and also provides a means to calibrate the forces
as described in the Appendix.
The scale bar is 200 $\mu$m, and the images have area fractions
$\phi=$ 0.88, 0.92, and 0.96.}
\label{fig:incline}
\end{figure}

We wish to use the droplet images (such as those shown in
Fig.~\ref{fig:incline}) to determine the forces droplets exert on
each other.  An isolated droplet is circular with 2D radius $r_0$
due to surface tension.  Droplets feeling forces from other droplets
are deformed.  Our goal is not to know the exact form of the force
law governing inter-droplet forces.
Rather, we need to know these forces to within
our experimental error.  The details of our approach are given in
the Appendix; we briefly summarize our method here.

By tilting the sample as shown in Fig.~\ref{fig:incline},
we exploit the known buoyant forces on the droplets, which are
$O(10^{-3}$~$\mu$N) per droplet.  Droplets further up the incline
feel larger forces and deform more because they must support
the buoyant weight of the droplets below.  This can be seen in
Fig.~\ref{fig:incline}, where droplets further up the incline are
more compressed.

The key parameters to determine the forces are the contact length $\Delta l$ of
the interface shared between the two droplets, and the modified
radius of curvature $r$ of each droplet deduced from the perimeters
of the droplets where they are not contacting another droplet.
Laplace's Law relates the interface curvature to the pressure
difference between the interior and exterior of a droplet; in 3D
this relation is $\Delta P = \gamma / r$ with $\gamma$ being the
surface tension.  From this, it is clear that deforming a droplet
(increasing its internal pressure) results in $r < r_0$.
Likewise, it is clear from Fig.~\ref{fig:incline} that $\Delta l$
is larger for droplets feeling larger forces (droplets farther up
the incline).

\begin{figure}[t]
\includegraphics[width=3.3in]{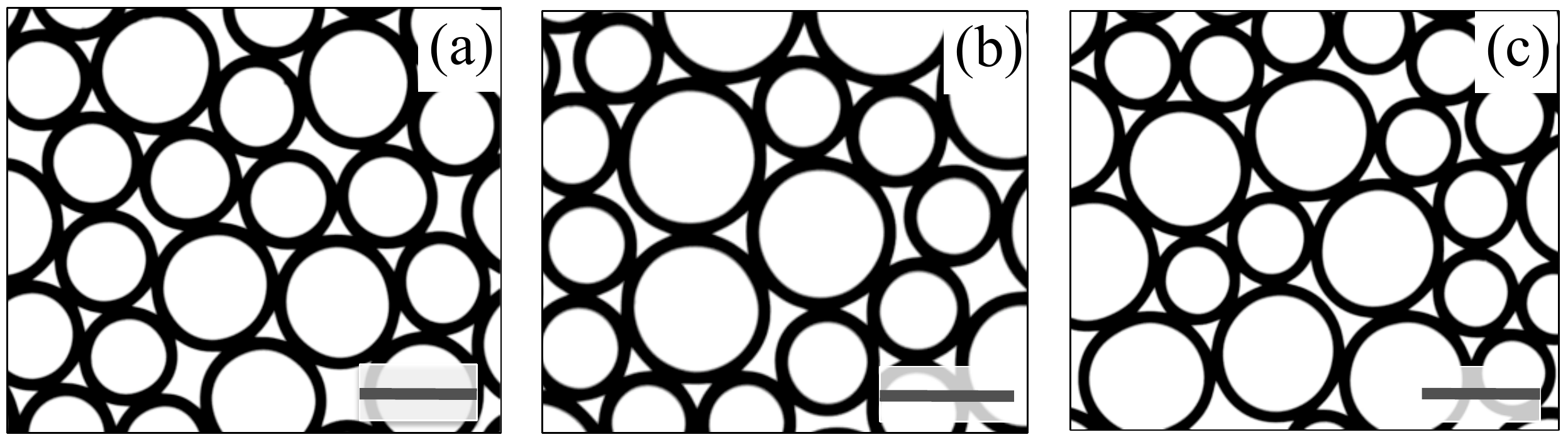}
\caption{Close up view of regions within our three bidisperse
samples. The scale bar in each image is 250 $\mu$m. (a) is an image
for our sample with size ratio 1.25 at $\phi = 0.89$. (b) is an
image for our sample with size ratio 1.42 at $\phi = 0.90$. (c)
is an image for our sample with size ratio 1.52 at $\phi = 0.90$.
}
\label{fig:BinarySample}
\end{figure}

To find a force law from our data, we use numeric methods to find a
force law $f(\Delta l, r)$ that best satisfies what we know about
the data:
(1) the net force on any droplet is
zero (as the droplets are motionless), (2) the components of
the forces in the $y$ direction balance the buoyant weight
of each droplet, and (3) the forces
acting between two touching droplets are equal and opposite.
After some work, we find for two droplets $i$ and $j$ in contact
that the best functional form is
\begin{equation}
F = \alpha_1 (R_0 L/r_{ij}) +\alpha_2 (R_0 L/r_{ij})^2, 
\label{forcelaw}
\end{equation}
where $1/r_{ij} = 1/r_i + 1/r_j$, $R_0$ is the mean 3D droplet
radius prior to putting the sample into the quasi-2D chamber,
and $\alpha_1$ and $\alpha_2$ are constants that depend on the
oil-water-surfactant details.  For our samples, $\alpha_1 = 4.25$
$\mu$N/mm and $\alpha_2 = 4.12$ $\mu$N/mm$^2$.  The forces we
observe in the experiment range from 0 - 0.5$\mu$N, several hundred
times larger than the buoyant weight of an individual droplet,
which makes sense given that the inter-droplet forces at high
$\phi$ are due to the accumulated pressure from the buoyant
forces of the many droplets below.  Intriguingly, we find that
the force law does not depend on the gap thickness of the sample
chamber.  We stress that Eqn.~\ref{forcelaw} is an empirical
deduction and is only approximate.
The data presented in the next section are from images taken with
the $1.6\times$ objective lens and the forces obtained have a random
uncertainty of 16\%.
For a more detailed discussion on
obtaining this force law and the uncertainty, see the Appendix.

\section{Jamming of Binary Packings}
\label{jamming}

We will investigate the jamming transition and force networks
in disordered bidisperse packings using the setup shown 
in Fig.~\ref{fig:incline}.  By allowing the sample to equilibrate
in this chamber, we set up a gentle gradient in area fraction
ranging from just barely jammed at the bottom of the incline 
to well-jammed near the top.  This lets us study the properties
of the packing over this full range of area fractions.  For the
analysis in this section, we only consider droplets between 5 and
50 droplet diameters up the incline, unless otherwise specified.
This region corresponds to an area fraction $\phi$ ranging from
$0.89-0.96$.  We study three different samples with different size
ratios, as shown in Fig.~\ref{fig:BinarySample}.  Details of each
sample are given in Table~\ref{table:Binary1}.

\begin{table}
\begin{center}
\begin{tabular}{cccccc}
\hline
$\sigma$ & $n_r$ & $r_{0}^{\rm (big)}$  & poly$^{\rm (big)}$ &
$r_{0}^{\rm (small)}$ & poly$^{\rm (small)}$ \\
\hline
1.25 & 0.684 & 126 $\mu$m & 3.4\% & 102 $\mu$m & 3.1\% \\
1.42 & 0.849 & 130 $\mu$m & 3.0\% & 105 $\mu$m & 3.4\% \\
1.52 & 0.806 & 137 $\mu$m & 3.3\% & 90.1 $\mu$m & 3.1\% \\
\hline
\end{tabular}
\end{center}
\caption{This table provides parameters characterizing the
droplets in our 3 bidisperse samples. The first column $\sigma =
r_{0}^{\rm (big)}/r_{0}^{\rm (small)}$ is the size ratio and the
second column is the number ratio $n_r=N^{\rm (big)} / N^{\rm
(small)}$.  The polydispersity in droplet sizes for the big
droplets is indicated by poly$^{\rm (big)}$ and for the small
droplets by poly$^{\rm (small)}$. The polydispersity is defined
as the standard deviation in droplet sizes normalized by the mean
size. Our uncertainties are $\pm$ 2 $\mu$m in $r_{0}$, $\pm$ 0.1\%
in $\sigma$, and $<$ 0.1\% in polydispersity. Since we image every
single droplet there is no measurement error in $n_r$.}
\label{table:Binary1}
\end{table}

\subsection{Identifying jamming area fraction $\phi_c$}

We start by identifying the jamming area fraction $\phi_c$ for
each data set. To determine the jamming point, we calculate
the area fraction with distance up the incline directly from
our images of the droplets in three steps. First, using the
centers and radii of the droplets we compute the radical Voronoi
cells~\cite{Aurenhammer87, Okabe2000} for each droplet.  The radical
Voronoi tessellation divides space into polygons, one per droplet,
taking into account each droplet's size so that each droplet is
fully contained within its own polygon.  Second, we determine the
area $A_v$ of each Voronoi cell and the area $A_d$ of each droplet.
Third, the area fraction $\phi(y)$ at a position $y$ is computed as
$\phi(y) = \Sigma{}_kA_{d,k}/\Sigma{}_kA_{v,k}$, where $k$ indexes
all droplets with a center of mass within $y - \Delta{}y/2$ and
$y + \Delta{}y/2$. For this step, all the droplets are examined,
down to nearly $y=0$, except for the droplets at $y \approx 0$
where the Voronoi cell is poorly defined. 
We choose $\Delta{}y = 6\langle r_{0} \rangle$ where $\langle r_0
\rangle$ is the mean droplet radius (which depends on the size
ratio and number ratio, see Table \ref{table:Binary1} for details).
This value of $\Delta y$ gives roughly 150 droplets per $y$ sampled.
Within this
window of $\Delta y$, $\Delta \phi = (\partial \phi/\partial y)
\Delta y \approx 0.007$.
From $\phi(y)$ we can obtain the jamming point $\phi_{c}$ by
extrapolating the value of $\phi$ to $y = 0$, where $y=0$ is
defined as the bottom of the droplet pile. We can treat the $y =
0$ point in our data as the jamming point since the forces between
droplets at $y = 0$ are nearly zero.  For the three data sets, the
extrapolation is done by fitting $\phi(y)$ to a power law [$\phi(y)
= \phi_c + a y^b$] giving $\phi_c = $0.855 $\pm$ 0.005, 0.861 $\pm$
0.005, and 0.858 $\pm$ 0.008 for the data with size ratio $\sigma$
= 1.25, 1.42, and 1.52, respectively. We chose to use $\phi(y)
= \phi_c + a y^b$ since $\phi - \phi_c$ vs $y$ appears fairly
linear on a log-log plot.  In simulations on frictionless disks and
experiments on 2D foams it has been reported that $\phi_c \sim 0.84$
for bidisperse systems~\cite{OHernPRL2002, Desmond2009, Katgert2010_2},
which is a little lower than the values we found.

Our measured area fraction depends on where
we define the outer perimeter of a droplet.   As seen in
Fig.~\ref{fig:BinarySample}, the droplets have thick black outlines.
We look at the outer edge of each outline, and define the perimeter
as the pixel location where the intensity is halfway between
the white color outside the droplet, and the black color in the
darkest part of the outline.  The transition from black to white
occurs over a distance of 2-3 pixels, and so we judge that we
have a systematic uncertainty in the area fraction of roughly 1\%
due to the determination of the perimeter position.  Since this
is systematic, the distance to the jamming point ($\phi-\phi_c$)
is insensitive to this error and therefore in most of our results
we focus on $\phi-\phi_c$.

\subsection{Critical Scaling}
\label{critscaling}

The first critical scaling we investigate is the coordination
number, the mean number of contacts each droplet has.  Prior
numerical studies of jamming in frictionless systems found
that the coordination number $z$ obeys a power law scaling of
the form $z - z_c = A(\phi - \phi_c)^{\beta_z}$, where $A \sim
3.5$, $z_c = 4$, and $\beta_z = 1/2$~\cite{Durian1995, Durian1997,
OHernPRL2002}. It has been observed that $A$ has a slight dependence
on the force law and polydispersity, but $z_c = 4$ and $\beta =
1/2$ are independent of the force law and polydispersity. Katgert
\textit{et al.}~\cite{Katgert2010_2} found for a 2D bidisperse foam
with size ratio 1.5 a critical scaling with $A = 4.02 \pm 0.02$
and $\beta_z = 0.50 \pm 0.02$ while fixing $z_c = 4$. The critical
point $z_c$ has been interpreted as the isostatic point $z_{iso}$
(minimum number of contacts necessary for a mechanically stable
packing). For 2D, $z_{iso} = 4$, in agreement with $z_c$ found in prior work.

\begin{figure}[t]
\begin{center}
\includegraphics[width=3.in]{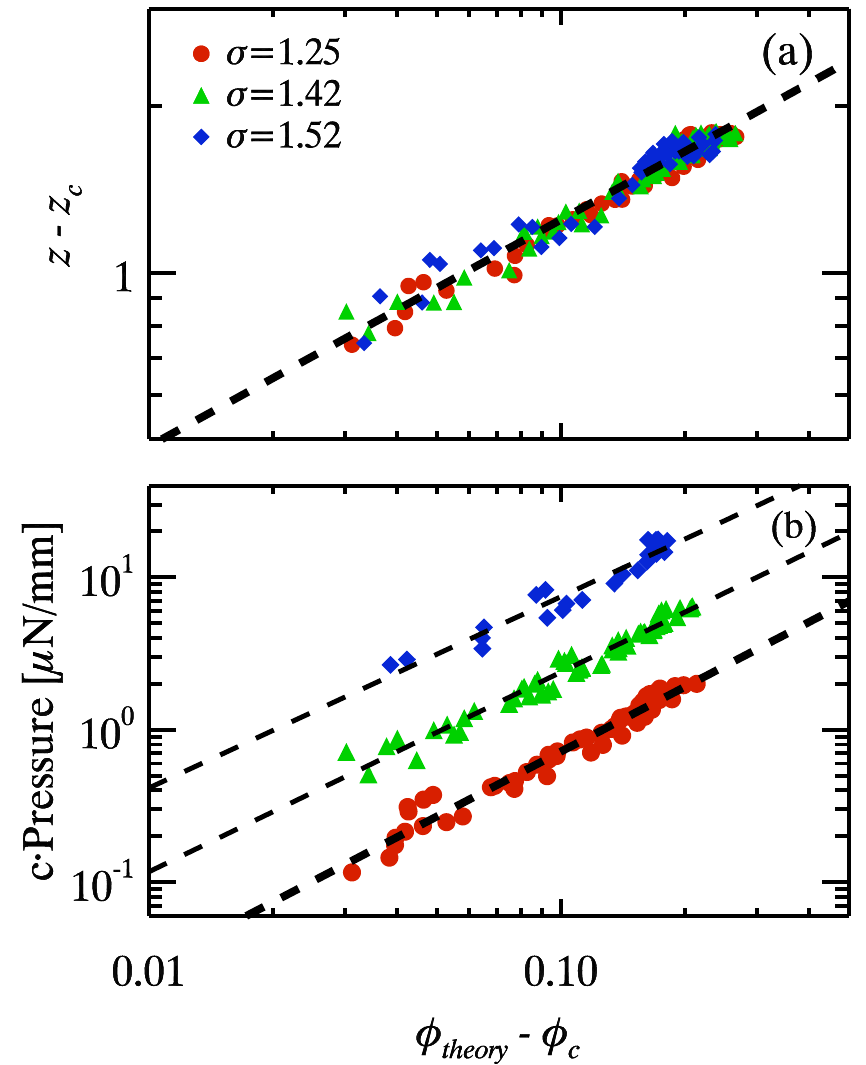}
\caption{(Color online) (a) Scatter plot of coordination number
against $\phi_{\rm theory} - \phi_c$.  All data were fitted
together to $z - z_c = A(\phi - \phi_c)^{\beta_z}$, where the fit
is shown as the black dashed line with fit parameters $z_c = 4.2$, $A = 3.2$, and $\beta_z = 0.4$. Fitting
the different data sets separately gives slightly different
fit values, listed in Table~\ref{table:FittingValues}. (b) A scatter plot
between pressure and $\phi_{\rm theory} - \phi_c$.  The pressure has
been scaled by $c = 1, \sqrt{10}$, and 10 for the $\sigma =
$ 1.25, 1.42, and 1.52 data respectively. Each data set is fitted
to $\mathcal{P} = A(\phi_{\rm theory} - \phi_c)^{\beta_P}$, 
shown as the black dashed lines. The fit values are
given in Table~\ref{table:FittingValues}.}
\label{fig:z_vs_phi}
\end{center}
\end{figure}


\begin{table}[t]
\begin{center}
\begin{tabular}{cccc}
\hline
 & \multicolumn{3}{c}{$z-z_c = A_{z}(\phi_{\rm theory} - \phi_{c})^{\beta_z}$} \\
$\sigma$ & $A_z$ & $\beta_{z}$ & $z_c$ \\
1.25 &  3.2 $\pm$ 0.6 & 0.4 $\pm$ 0.2 & 4.3 $\pm$ 0.3 \\
1.42 &  3.3 $\pm$ 0.6 & 0.4 $\pm$ 0.2 & 4.3 $\pm$ 0.3 \\
1.52 &  3.2 $\pm$ 0.7 & 0.3 $\pm$ 0.2 & 4.0 $\pm$ 0.4 \\
\hline
\hline
 & \multicolumn{2}{c}{$\mathcal{P} = A_P(\phi_{\rm theory} - \phi_c)^{\beta_P}$} & \\
$\sigma$ & $A_{P}$ [$\mu$N/mm] & $\beta_{P}$ &  \\
1.25 & 19 $\pm$ 1 & 1.41 $\pm$ 0.03 &  \\
1.42 & 15 $\pm$ 1 & 1.30 $\pm$ 0.03 &  \\
1.52 & 13 $\pm$ 2 & 1.26 $\pm$ 0.07 &  \\
\hline
\hline
 & \multicolumn{2}{c}{$f_{ij} = F_{0}(\delta{}r_{ij}/d_{ij})^{\beta_f}$}  & \\
$\sigma$ & $F_{0}$ [$\mu$N] & $\beta_{f}$ & \\
1.25 & 2.3 $\pm$ 0.2 & 1.27 $\pm$ 0.03 &  \\
1.42 & 2.4 $\pm$ 0.1 & 1.19 $\pm$ 0.02 &  \\
1.52 & 2.0 $\pm$ 0.1 & 1.15 $\pm$ 0.03 &  \\
\hline
\end{tabular}
\end{center}
\caption{The fitting parameters for the power law fits to the
data for each size ratio $\sigma$.
Note that
simulations found $\beta_P = \beta_f$ \cite{OHern2003}; see
text for a discussion.  The uncertainties in the fit values are
obtained by computing the standard error in each fitting parameter.}
\label{table:FittingValues}
\end{table}%

To compare experimental data and simulation data, the experimental
area fraction needs to be converted into a theoretical area fraction
\cite{Katgert2010_2}.  This is because the simulated particles
are allowed to overlap (thus diminishing the total area they
take up at large $\phi$) while our experimental droplets always
occupy the same total area.  We convert our experimental $\phi$
values to $\phi_{\rm theory}$ using the method of Katgert {\it
et al.}~\cite{Katgert2010_2}.  From our data we determine $z$
and $\phi_{\rm theory}$ at various points along the incline.
The results are plotted relative to the jamming point in
Fig.~\ref{fig:z_vs_phi}(a), and show power-law scaling. Fitting
the each data set to the theoretical scaling law, $z - z_c =
A(\phi - \phi_c)^{\beta_z}$, we obtain values for $A$, $z_c$, and
$\beta_z$ which are reported in Table~\ref{table:FittingValues}.
Our values of $A \approx 3.2$ are close to $A \approx 3.5$ found in a numerical study by O'Hern \textit{et al.}~\cite{
OHernPRL2002} for particles with size ratio 1.4. The fitted values for $z_c$
are within the uncertainty of the previously found value of
4~\cite{Durian1995, Durian1997, OHernPRL2002, Katgert2010_2}.
However, our droplets have a slight attraction which may result
in a slightly tighter packing of droplets at $\phi_c$ 
with a coordination number $z_c> 4$. 
Given our uncertainties of
$z_c$, our data are consistent with both $z_c=4$ and $z_c>4$.
Finally, for each packing, the exponent $\beta_z \approx 0.4$
agrees with the prior findings ($\beta = 0.5$) to within our
uncertainty, although we have a fairly large uncertainty in our
exponents. Interestingly, in 2D photoelastic disk experiments,
they found $z - z_c = (\phi - \phi_c)^{\beta_z}$ with $\beta_z =
0.53 \pm 0.03$ without needing to convert their experimental $\phi$
to $\phi_{\rm theory}$~\cite{Majmudar07}, but $A \sim 25$ for that
study which is considerably different from our results.
In their work, they were limited to area fractions close to $\phi_c$
due to the difficulty of compressing their particles to high area
fractions, while our data (and those of Ref.~\cite{Katgert2010_2})
extend over a larger range of $\phi$.

The second critical scaling we investigate is the dependence
of pressure $\mathcal{P}$ with distance to the jamming point.
Simulations of 2D particles found $\mathcal{P}
= A (\phi_{\rm theory} - \phi_{c})^{\beta_P}$, where $A$ and
$\beta_P$ depend on the form of the force law.  In the numerical
study by O'Hern \textit{et al.}~\cite{OHern2003}, they used
frictionless disks that interacted via the force law $f_{ij} =
F_{0}(\delta{}r_{ij}/d_{ij})^{\beta_f}$, where $F_{0}$ is a scale,
$\delta{}r_{ij}$ is the distance between two particles in contact,
and $d_{ij}$ is the sum of the radii of the particles in contact.
They found that $\beta_{P} = \beta_f$.  It is
certainly possible that for other force laws, the scaling of
pressure with $(\phi-\phi_c$) could be different.  In particular,
in our experiment, the force between two droplets is not a
unique function of $\delta r_{ij}$ but rather depends on the
droplet perimeters which are influenced by all of their neighbors.
In 2D photoelastic disk experiments $\beta_P$
was found to be 1.1~\cite{Majmudar07}.  No prior experimental 2D
studies have examined the scaling of $\mathcal{P}$ for systems
without static friction.

For our experiment, we compute the local pressure of our
sample by first locating a set of droplets $k$ within a
window $y - \Delta{}y/2$ and $y + \Delta{}y/2$.  For these
$k$ droplets the pressure is $\mathcal{P} = \sum_{i}\sum_{j >
i}F_{ij}r_{ij}/\Sigma{}_{k}A_v$, where $i$ and $j$ index all
contacts on the $k$ droplets and $\Sigma{}_{k}A_{k,v}$ is the
sum of the Voronoi areas of all $k$ droplets~\cite{Allen89,
OHern2003}. In this formula, $F_{ij}$ and $r_{ij}$ are both taken
to be positive scalars.  Here we use $\Delta{}y = 5r_{0}$. In
Fig.~\ref{fig:z_vs_phi}(b) we plot the pressure for all three
packings against $\phi_{\rm theory} - \phi_c$. These results show
power-law scaling.  The dashed lines are the fit to $\mathcal{P}
= A (\phi_{\rm theory} - \phi_{c})^{\beta_P}$ with the fit values
shown in Table~\ref{table:FittingValues}.  In particular, we find
$\beta_P$ values between 1.26 - 1.41, larger than $\beta_P = 1.1$
found for photoelastic disks \cite{Majmudar07}.

\begin{figure}[t]
\includegraphics[width=3.1in]{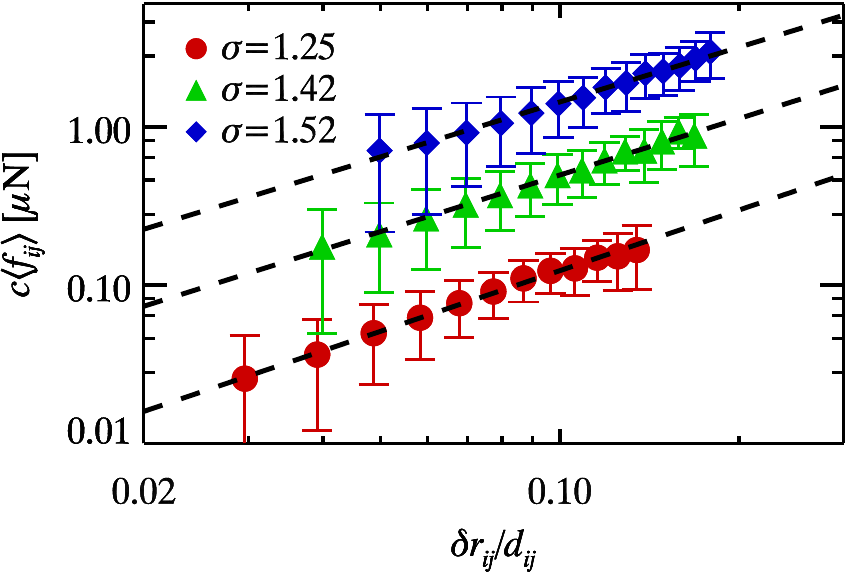}
\caption{(Color online) The average force between droplets in
contact plotted against the amount of compression between the
droplets. The average force has been scaled by a prefactor of $c
= 1, \sqrt{10}$, and 10 for the $\sigma = $1.25, 1.42, and 1.52 data
respectively. Each data is fitted to $\langle{}f_{ij}\rangle{} =
F_0 (\delta{}r_{ij}/d_{ij})^{\beta_f}$ and the fits are
shown as the black dashed lines. The fit values are given in
Table~\ref{table:FittingValues}. Note that this data is an effective
force law, not the true force law: for a given
$\delta{}r_{ij}/d_{ij}$,
different droplet pairs may experience different contact forces.
To illustrate this, we have added error bars to the plot, where
the error bars represent one standard deviation in the spread of
measured contact forces at each $\delta{}r_{ij}/d_{ij}$.
}
\label{fig:force_vs_compression}
\end{figure}

To compare with the simulations of O'Hern \textit{et
al.}~\cite{OHern2003}, we wish to approximate how forces between
our droplets depend on their separations $\delta r_{ij}$.  For each
observed $\delta_{ij}$ we find the true force $f_{ij}$ from our
force law.  We average all of the observations over small windows in
$\delta_{ij}$ to find an effective average force law as a function
of $\delta_{ij}$, plotted in Fig.~\ref{fig:force_vs_compression}.
The error bars emphasize that Fig.~\ref{fig:force_vs_compression} is
only an average trend rather than the true force law.  Intriguingly,
the averaged data follow a power law:  we fit each data to
$\langle{}f_{ij}\rangle{} = F_{0}(\delta{}r_{ij}/d_{ij})^{\beta_f}$
to obtain the power law exponent $\beta_f$.  The fits are shown
as the black dashed lines in the figure, with fit values listed
in Table~\ref{table:FittingValues}.

Our fits give $\beta_f < \beta_P$ 
in contrast to the results of O'Hern \textit{et
al.}~\cite{OHern2003} where $\beta_P = \beta_f$.  This equality was
found for systems close to the jamming area fraction.  The exponent
for the pressure, $\beta_P$, relates to how droplets are compacted
with increasing $\phi - \phi_c$~\cite{Ellenbroek2006}.  Close to
$\phi_c$, when $\phi$ is slightly increased, droplets can avoid
significant compression by rearranging and forming more contacts,
however, at larger $\phi$, droplets can not form many new contacts
and must instead undergo larger compression. Therefore, at larger
area fractions, the pressure increases more rapidly with $\phi
- \phi_c$ than it does near the jamming point.  This argument
predicts $\beta_{P} > \beta_f$, in agreement with our data which
extends far from $\phi_c$.  
While the uncertainty in each force measurement is 16\%, this uncertainty is unlikely
to significantly affect the pressure results, as the data of
Figs.~\ref{fig:z_vs_phi} and \ref{fig:force_vs_compression} are averages over many forces.

\subsection{Force Distribution}
\label{forcedistribution}

\begin{figure}[t]
\includegraphics[width=3.1in]{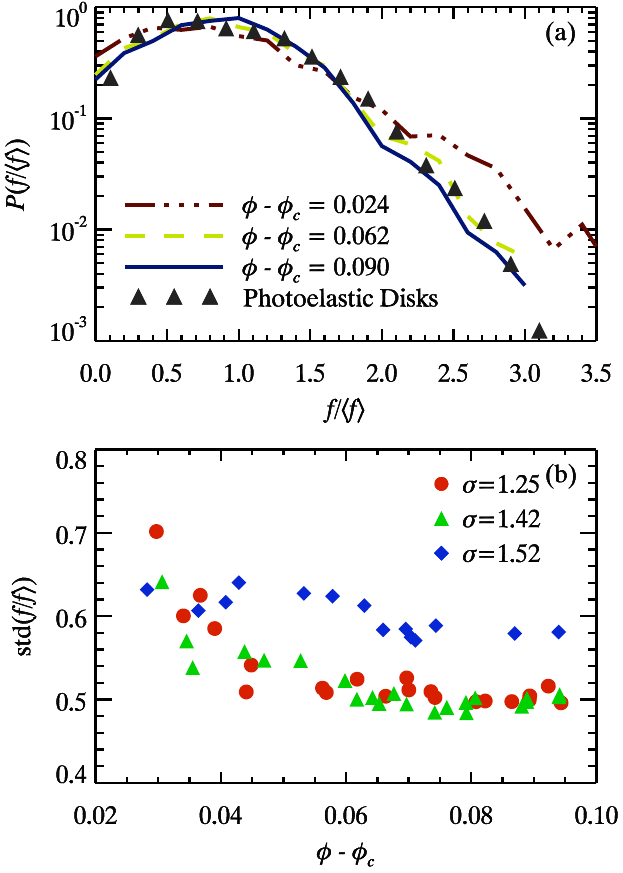}
\caption{(Color online) (a) Distribution of contact forces relative
to the mean contact force at different $\phi$ for the $\sigma =
1.25$ packing.  The mean force is $\langle{}f\rangle{} = 0.011$
$\mu$N, 0.045 $\mu$N, and 0.13 $\mu$N for the $\phi-\phi_c = 0.024,
0.062$, and 0.106 data respectively. The solid triangles are data from
a 2D photoelastic disk experiment at $\phi - \phi_c \sim 0.016$
from Majmudar \textit{et al.}\cite{Majmudar05}.
(b) The standard deviation of
$P(f/\langle{}f\rangle{})$ for each packing at different $\phi$.
The standard deviation of the Majmudar \textit{et al.} data is 0.52.
}
\label{fig:p_of_f}
\end{figure}

We now consider the distribution of contact forces for each packing
at different area fractions. Like before, we sample the contacts
forces at various points up the incline using a window of $y -
\Delta{}y/2$ and $y + \Delta{}y/2$. However, we need many contacts
to obtain a good distribution of contact forces, and therefore,
we use $\Delta{}y = 30r_{0}$.  (Over this range of $\Delta{}y$
and for droplets at least 10 diameters up the incline, $\Delta
\phi = (\partial \phi / \partial y)\Delta y = 0.025$.)
This window size gives roughly
2500 contacts for each $y$ sampled. In Fig.~\ref{fig:p_of_f}(a),
the lines show the distribution of contact forces normalized by the
mean contact force at locations with $\phi-\phi_c$ as indicated;
all the data are for the $\sigma = 1.25$ packing.  In all our data
most forces are near or less than the mean force $\langle{}f\rangle$
and that the maximum force is about $3\langle{}f\rangle$, with
a somewhat exponential tail.  The shape and magnitude of all the
curves are roughly the same.  All curves show a dip at small forces.
The symbols in Fig.~\ref{fig:p_of_f}(a) show the distribution of
normal forces from Majmudar \textit{et al.}~\cite{Majmudar05},
an experiment using frictional 2D photoelastic disks. In their
experiment, the particles were isotropically compressed to an
area fraction $\sim$ 0.016 above the critical area fraction.
Our results look essentially the same as theirs, despite the
differences in experimental systems.  These distributions are also
similar to simulations and experiments on 2D and 3D frictionless
systems on foams and emulsions~\cite{OHernPRL2002, Katgert2010_2,
Brujic2003_2, Zhou2006}. This suggest that the distribution of
forces is a fairly universal property of all jammed
systems. The shape of our measured distribution is also in
general agreement with the predictions of the force network
ensemble~\cite{Snoeijer2004} and is also in agreement with the
earlier q-model~\cite{Coppersmith1996}, although our data are not
clear enough to distinguish between these two models.

As observed by Katgert \textit{et al.} \cite{Katgert2010_2},
our force distributions slightly narrow with increasing
area fraction in Fig.~\ref{fig:p_of_f}(a). To quantify this,
we plot the standard deviation of each force distribution in
Fig.~\ref{fig:p_of_f}(b). For the $\sigma = 1.25$ and $\sigma =
1.42$ packings, the width of the force distributions decreases
quickly above the jamming point and then saturates to a constant
width of $\sim$ 0.5.  The $\sigma = 1.52$ packing has a broader
distribution of forces at larger $\phi - \phi_c$ compared to the
other two packings, and the decrease in the width as $\phi - \phi_c$
increases is more subtle.  Overall, our results are qualitatively
in agreement with \cite{Katgert2010_2}, although they did not
compute the standard deviations.  
It is possible that measurement errors in our forces
have a dependence on $\phi$, as discussed in the Appendix, and that
this could affect the dependence of the standard deviation on
$\phi$.  However, the errors are no worse than 16\%, much less than
the width of the distributions for any $\phi$, and so a
possible $\phi$-dependence of the errors cannot account for the decrease in width seen in
Fig.~\ref{fig:p_of_f}(b).

\subsection{Force Chains}
\label{forcechains}

\begin{figure}
\includegraphics[width=3.3in]{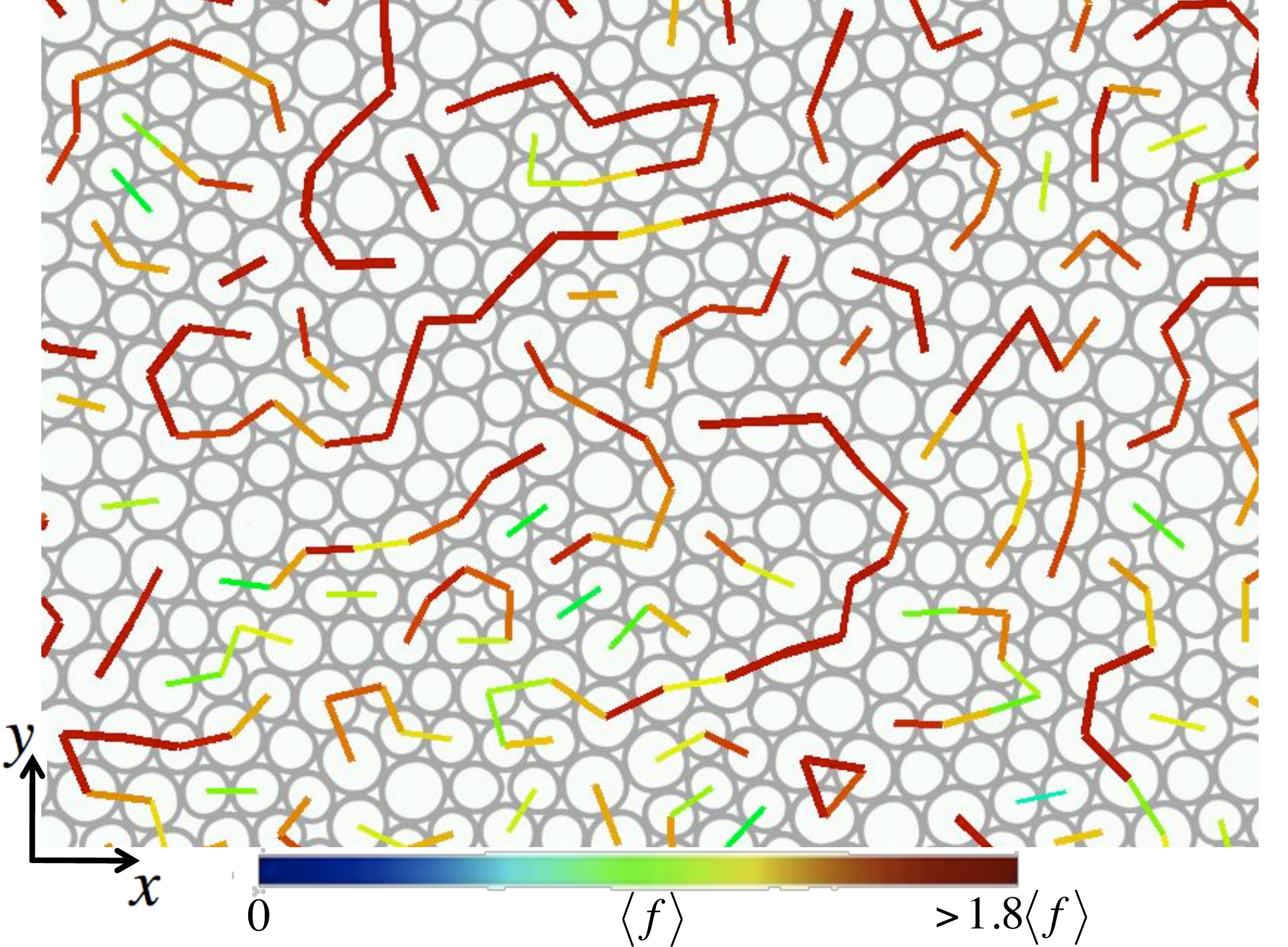}
\caption{(Color online). This image shows only the forces
belonging to a force chain within a region of the $\sigma =
1.25$ sample. On average, the forces
are larger further up the image because the sample is inclined,
and this can be observed in the image by the increasing redness
of the force segments at the top.}
\label{fig:BigImage}
\end{figure}

In this section we consider various statistical measurements on
the randomness of the force chain network, and then we compare
predictions of the Bruji\'{c}-Zhou model~\cite{Brujic2003_2,
Brujic2003, Zhou2006, Zhou2009} to our data. When analyzing data
in this section we consider all droplets and contacts between $40
\le y/2r_o \le 80$, and over this range $\phi$ increases from 0.93
to 0.96. We find that all of the properties discussed below do not
depend on $\phi$ at larger area fractions, and so considering this
larger range of $\phi$ gives us better statistics.

To start, we define a force segment to belong to a force chain if
it is one of the two largest forces on both droplets joined by the
force segment. Under this definition, each droplet can only have
a maximum of two force segments that belong to a force chain,
and therefore, our definition does not allow for force chain
branching or merging, which will simplify the analysis below.
Note that even droplets with small forces can participate in force
chains, given that the definition only requires the force segments
to be large for the given droplet and its contacting neighbors.
Figure~\ref{fig:BigImage} shows force chains using our definition.
These chains are fairly linear and vary in length from 1 to more
than 10 force segments.  Our samples are formed by slow compaction
due to buoyant forces, which may introduce an anisotropy in the
force chain network~\cite{Cates1998, Majmudar05, Toyia_2004}.
Indeed, we find a slight bias for force chain segments to be orientated
up the incline.


\begin{figure}[t]
\begin{center}
\includegraphics[width=1.3in]{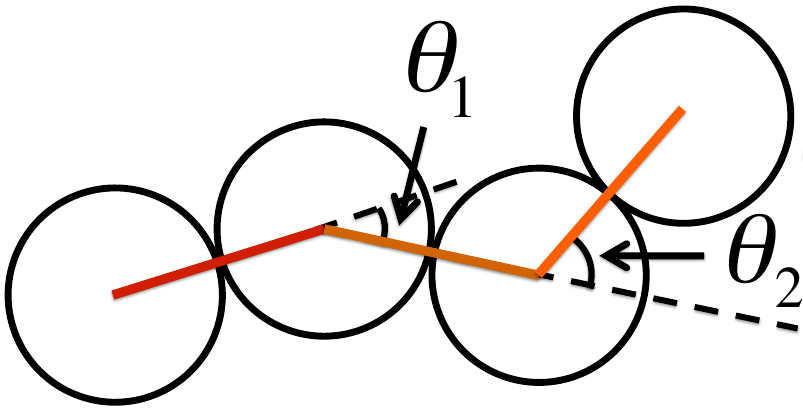}
\end{center}
\caption{(Color online). Definitions of the angles $\theta_1$
and $\theta_2$ between joining force segment. In the sketch both
$\theta_1$ (clockwise to extended line) and $\theta_2$ (counter clockwise to extended line) are positive. If there is a correlation in orientation that tends to make force chains linear,
then the correlation between $\theta_1$ and $\theta_2$ is positive.}
\label{fig:correlation}
\end{figure}

Given that force chains form linear like structures and that
there is a slight anisotropy for force chains to align along the
incline we may expect there to be correlations in the orientation
of neighboring force chain segments. To quantify such a correlation
we define two relative angles $\theta_1$ and $\theta_2$ between
joining force segments, where the definition of $\theta_1$ and
$\theta_2$ are shown in Fig.~\ref{fig:correlation}.  We compute
the Pearson correlation coefficient $C = \mbox{cov}(\theta_1,
\theta_2)/\sigma_{\theta}^2$, where $\mbox{cov}(\theta_1,
\theta_2)$ is the covariance of $\theta_1$ and $\theta_2$
and $\sigma_{\theta}$ is the standard deviation of $\theta$.
We find that for all $\phi$, $C$ is zero or nearly zero (at most
$C = 0.2$), indicating no correlation.  This agrees with prior
work on 3D emulsions~\cite{Brujic2003_2, Zhou2006}.  Thus,
the apparent linearity of force chains seen in some locations of
Fig.~\ref{fig:BigImage} is not due to correlations in the relative direction of neighboring segments that would keep
the chain straight.

\begin{figure}
\includegraphics[width=3.3in]{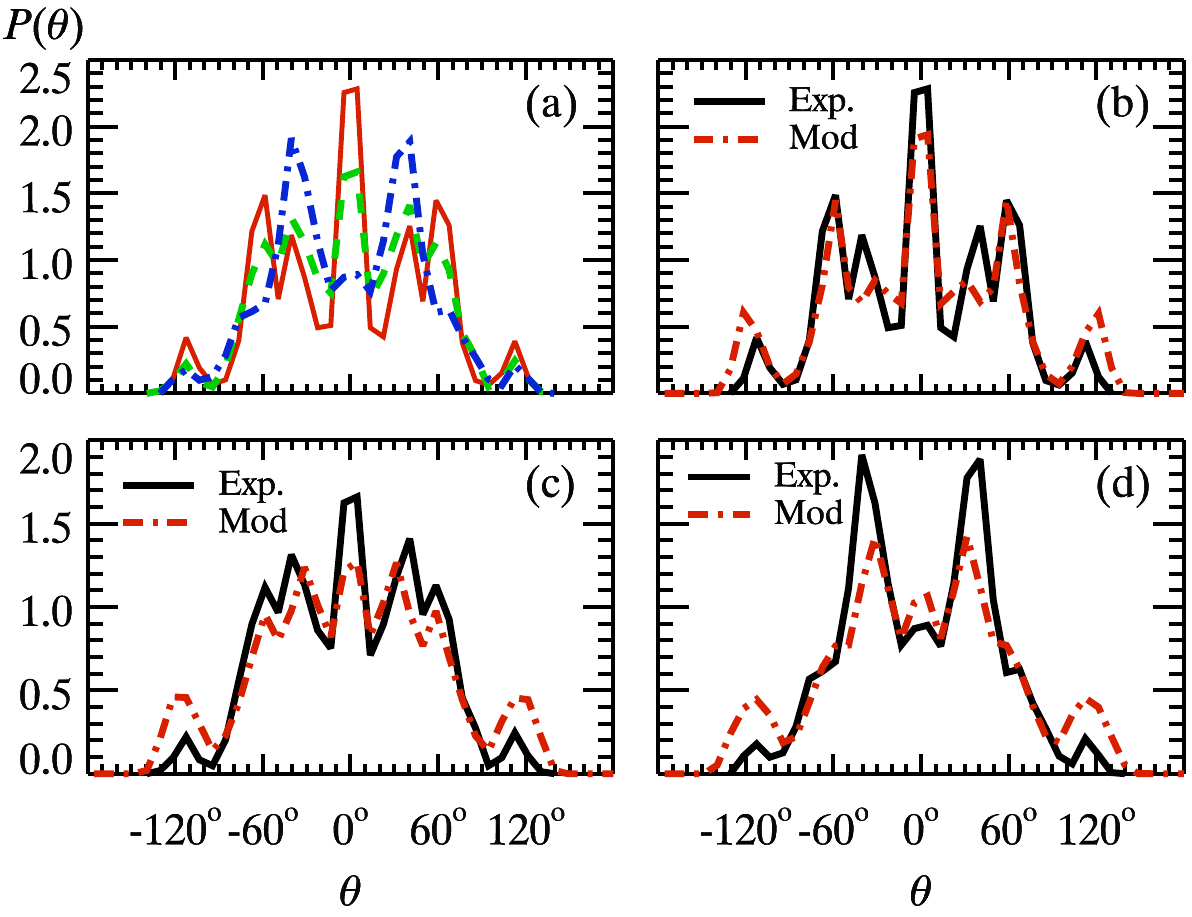}
\caption{(Color online) (a) Distribution of $\theta$ for each
packing, where both $\theta_1$ and $\theta_2$ are treated as a
single variable $\theta$. The red solid line is the distribution
for the $\sigma = 1.25$ packing, the green dashed line is the
distribution for the $\sigma = 1.42$ packing, and the blue
dashed-dot line is the distribution for the $\sigma = 1.52$
packing. (b-d) Comparisons between the experimental distributions
and the predictions of the
Bruji\'{c}-Zhou model, for size ratios
(b) $\sigma=1.25$, (c) $\sigma=1.42$, and (d) $\sigma=1.52$.
}
\label{fig:theta_distribution}
\end{figure}

To further explore the tendency for force chains to be linear,
we consider the distribution of $\theta_1$, where we drop the
subscript $1$ as we are only focusing on two force segments at a
time rather than three.  In Fig.~\ref{fig:theta_distribution}(a)
we plot the distribution in $\theta$ for all three packings. The
distribution shows that most force chain segments form at an angle
$|\theta{}| < 60^{\circ{}}$.  
Thus, force segments tend to form a linear chain not because their
orientations are correlated, but simply because it's more probable
that they are oriented at small angles relative to each other.
Using our $P(\theta)$ data, we determine the
persistence lengths $l$ using the standard definition of persistence length for polymer chains.

We find 
$l=4.4\langle r_0
\rangle$, $4.8 \langle r_0 \rangle$, and $3.8 \langle r_0
\rangle$ for the $\sigma=1.25, 1.42,$ and 1.52 data.  These are the
distances beyond which the force chain has ``forgotten'' its
original direction. In analyzing the distributions similar to $P(\theta)$ for 3D emulsions, Zhou \textit{et al.}~\cite{Zhou2009} found a persistence length slight larger around $l \sim 6 -8 \langle r_0 \rangle$.

To further consider the orientations of force segments in force
chains, we consider a model proposed by
Bruji\'{c} \textit{et al.}~\cite{Brujic2003_2, Brujic2003} and
extended by Zhou \textit{et al.}~\cite{Zhou2006, Zhou2009}. The
Bruji\'{c}-Zhou model is a method for generating ensembles of local
particle configurations (a central particle and contacting first
neighbors) and the forces acting on a central particle by its first
neighbors. Each local configuration is generated by randomly placing
$z_{i}$ contacting neighbors such that any two neighboring particles
do not overlap. Next, the contact forces between the central
particle and $z_i - 2$ neighboring particles are chosen at random
from a distribution $P(f)$, leaving two unknown contact forces. We
choose $P(f)$ to match our experimentally measured distributions
(see Fig.~\ref{fig:p_of_f}). By invoking force balance, the two
remaining contact forces are found algebraically. Once a sufficient
number of local configurations are generated, the distribution of
force chain orientations can be studied. The basic assumptions
of this model are force balance, randomness in the magnitude of
forces, and randomness in the orientation of forces. For our data
the first assumption applies because the system is in mechanical
equilibrium and above we have shown that the other two assumption
reasonably apply.

One issue in using the Bruji\'{c}-Zhou model to predict
$P(\theta)$ is that the model only gives the forces between a
central droplet and its first neighbors. To define a force
chain segment we also need to know all the forces acting on each
first neighbor as well.  We therefore extend their model by
generating additional forces on the neighboring droplets in
exactly the same way (constrained by the forces already chosen
for the central droplet).  This lets us apply our force chain
definition given above, which requires that force segments be
among the largest two forces on both droplets the force acts
between.  We repeat this extended Bruji\'{c}-Zhou algorithm many
times to compile data from all cases where the algorithm gives
an instance of two valid force segments so that we can determine
$\theta_1$.
To make the inputs into the model as consistent as possible
with our experimental data, instead of randomly generating local
configurations, we randomly select local configurations from our
experimental data.

Figure \ref{fig:theta_distribution}(b)-(d) compares $P(\theta)$
measured in our experiments (black solid curves) with the
predictions of the model (red dot-dashed curves).  The model
is in good agreement with the experiment, with the exception
of some discrepancies in the magnitudes of the peaks.
The model captures significant features of the data:
for instance, the peak around $\theta = 0^\circ$ is much
different between Fig.~\ref{fig:theta_distribution}(b) and
Fig.~\ref{fig:theta_distribution}(d), and the model replicates
this difference.  We also note that if we loosen the
definition of force chain segments to simply those forces that
are the largest two forces acting on any droplet (independent of
how large they are relative to forces on neighboring droplets),
we find nearly identical distributions as the ones shown in
Fig.~\ref{fig:theta_distribution}.

\begin{figure}
\includegraphics[width=3.3in]{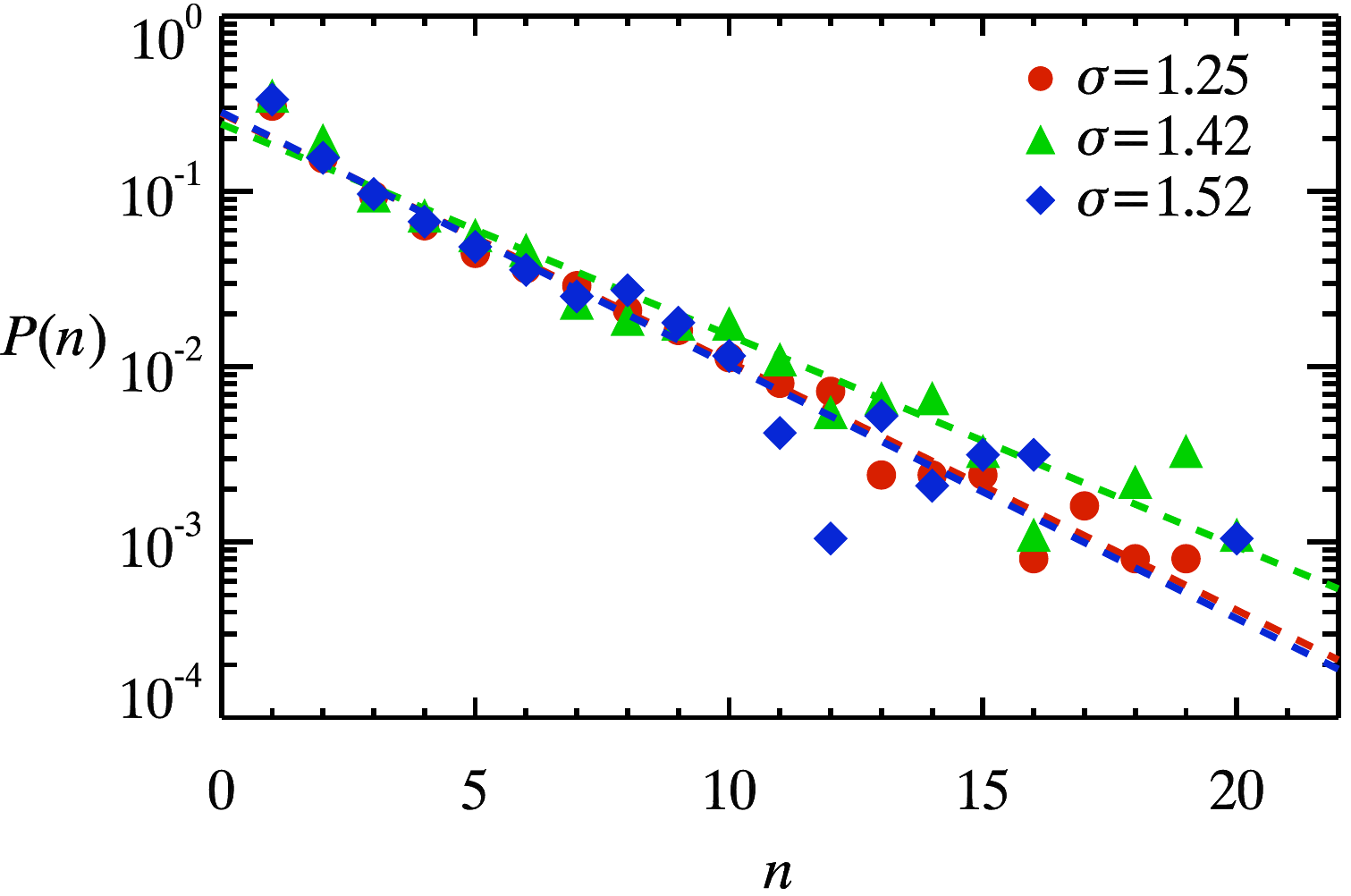}
\caption{(Color online)
Distribution of the number of force segments making up distinct
force chains. The data points are experimental values
and the dashed lines are fits to the data of the form $P(n) =
(1-p)p^n$, where $p$ is found to be 0.722, 0.758, and 0.717 for
the $\sigma =$ 1.25, 1.42, and 1.52 packings, respectively.}
\label{fig:chain_lengths}
\end{figure}

Our analysis suggests so far that the force chain network is random,
without long-range correlations.  It therefore seems plausible
that the distribution of force chain lengths should obey a random
process.  If there is a probability $p$ for a force chain segment
to be connected to a neighboring force chain segment, then the
distribution of chain lengths should obey the scaling $P(n) =
(1-p)p^n$, where $n$ is the number of force segments
within a force chain.  In
Fig.~\ref{fig:chain_lengths} we plot the distribution of chain
lengths for each packing. The data decay exponentially over
3 orders of magnitude.  The data are fit by $P(n) = (1-p)p^n$
with $p \approx 0.73$ (see caption for details), indicating that
it is highly likely that for a force chain to propagate through
the material.  The fits are shown as the dashed lines and show
good agreement with the data other than at $n=1$.

In granular quasi-static intruder simulations with friction between
particles by Peters \textit{et al.}, using a more sophisticated
definition of force chains, they also found an exponential
distribution of chain lengths~\cite{Peters2005}. From their reported
data on $P(n)$, we estimate a value of $p = 0.65$.  It appears
that statistically a force chain can be thought of as a random
process with probability $p$ for the force chain to propagate,
independent of $\phi$ but perhaps depending on the sample details.

\section{Conclusions}

We have introduced a new experimental model system composed of
quasi-2D emulsions droplets to study the jamming transition. Our
droplets are circular in shape and deform when press into one
another, and at the contacts between two droplets the forces are
in-plane mimicking a true 2D system. We can accurately measure
the forces between touching droplets to within 8\%, where our
method is not limited to our experiment, and could be extended
to determine forces in 2D foams, 3D emulsions, and 3D foams. Our
model system has unique strengths; we can easily make samples
with any distribution in particle sizes, emulsions are stable over
many days, setup is cheap, our droplets have no static friction,
and our method can be extended to cases of flow~\cite{Chen2012}.

Using our model system we observed power-law scaling of the
contact number and pressure with $\phi - \phi_c$, similar to
prior numerical models~\cite{Durian1995, Durian1997, OHernPRL2002,
OHern2003}.  Notably we
find that all three fit parameters for the contact number scaling
are quite close to the values found in 2D simulations. We verify
experimentally for the first time a link between the interparticle
force law and the critical pressure exponent, illustrating a direct
relationship between the bulk properties of an amorphous solid and
the interaction between the constituent particles. The agreement of
our results and the numerical models shows that the qualitatively
different particle interaction we have does not play a significant
role in determining the geometric structure and bulk modulus.

Our analysis of the inter-particle forces found a probability
distribution of forces in good agreement with those found in
prior experiments and simulations, strongly suggesting that the
shape is universal.  We further examined the spatial structure of
the large forces (``force chains'').  The directions of
neighboring force chain segments are uncorrelated although there
is a tendency for two force chains to be in the same direction.
This is a sensible result as this allows the large forces acting
on a droplet to balance one another.  The Bruji\'{c}-Zhou model,
which assumes random and uncorrelated force segments, recovers our
experimentally observed probability distribution of angles between
adjacent force segments.

This work provides more evidence for the universality of various
properties of the jamming transition, such as critical scaling,
the shape of the force distribution, and the structure of the
force network.


\section*{Acknowledgments}

We thank G.~Hunter and G.~Hentschel for helpful discussions.
This work was supported by the donors of The Petroleum Research
Fund, administered by the American Chemical Society (grant
47970-AC9), and additionally by the National Science Foundation
(grant CBET-0853837).

\appendix
\section{Method For Determining Force Law}

In this section we describe in detail our method for determining
an empirical force law that relates the outline of droplets
to the contact forces. For an overview of our method see
Sec.~\ref{sec:EmpiricalForceLaw}.  This section is organized
in the following manner: first, we discuss the measurements from
droplet outlines; second, we discuss the general form of possible
force laws; third, we present the optimization problem; fourth, we
deduce the best force law consistent with the data.  

\subsection{Measurable Variables}
\label{measurablevariables}

In this subsection, we discuss the various quantities measurable
from droplet images, and their measurement errors.  In the
following subsections, these quantities will be used to determine
the forces between droplet pairs.

\begin{figure}[t]
\includegraphics[width=3.3in]{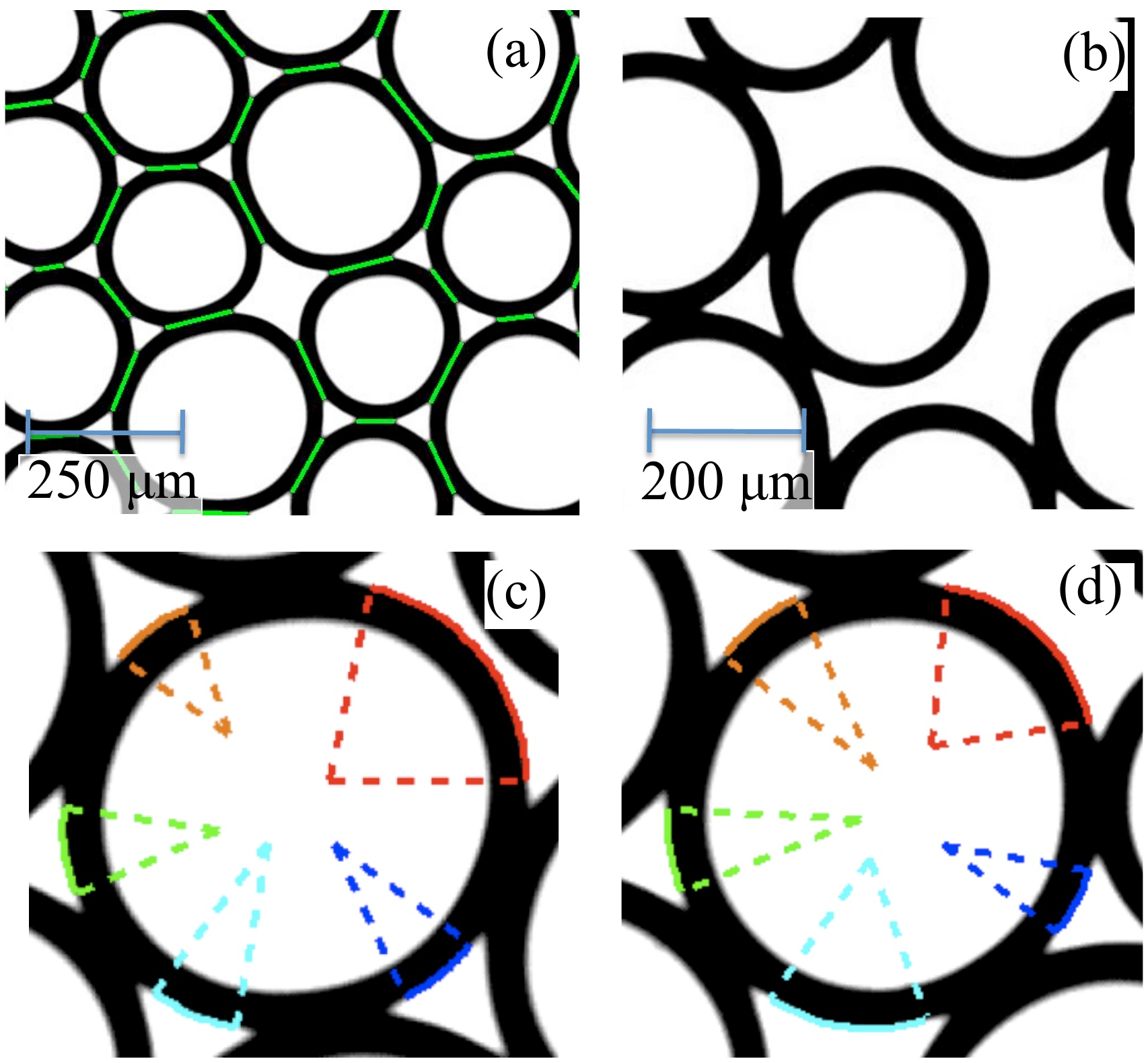}
\caption{(Color online). (a) Experimental image of droplets with the contacts found using our algorithm indicated by green lines. (b) Experimental image of a rattler droplet being held in contact with two neighboring droplets due to a slight adhesion. In this image the droplet are motionless and the system is not inclined. (c, d) A close up view of a droplet. Each portion of the water-oil interface is fitted to an arc with constant radius of curvature. The fits are shown as the different colored arcs.}
\label{fig:rattler}
\end{figure}

The larger the contact between two droplets, the more force they
feel.  This is quantified by the contact length $l_{ij}$ between
two droplets.  We measure this by identifying the portion of each
droplet's perimeter that is shared between them, shown
as the light green lines in Fig.~\ref{fig:rattler}(a). $l_{ij}$
is calculated as the length of the line segment. Since we can only
measure the two endpoints of each contact to 1 pixel accuracy,
we have an uncertainty $\delta{}l$ of $\sqrt{2}$ pixels. For our
highest magnification lens ($5\times$) this gives $\delta{}l =
1.1$ $\mu$m and for our lower magnification lens ($1.6\times$)
this gives $\delta{}l = 1.96$ $\mu$m.

One expects that any two droplets with a nonzero contact length
($l_{ij} > 0$) would experience a repulsive force at the contact.
However, we observe a slight attractive interaction between droplets
as shown in Fig.~\ref{fig:rattler}(b), where the central droplet
is adhering to two neighboring droplets. In this particular
configuration, the adhesion force is balanced by the repulsive
force, and therefore, the net force at each contact on the
central droplet is zero for some finite contact length $l_{0}$.

We determine $l_{0}$ by averaging together the contact lengths
found between 25-50 droplets at the very bottom of the incline.
The standard deviation of these contact lengths is about 0.025$r_0$,
which is one to two orders of magnitude smaller than measured
contact lengths $l_{ij}$ of deformed droplets in contact, and
therefore we conclude that $l_0$ is reasonably well-defined by
the mean value.  In addition to the adhesion, some of the finite
contact length is due to optical resolution limits resulting in a
systematic effect:  for similar conditions, $l_0$ is shorter when
using the $1.6 \times$ lens (data in Table 3) as compared to the
$5\times$ lens (data in Table 4).  Accordingly, the force law will
be taken to depend on $\Delta l = l - l_0$, canceling the systematic
effect, and the force law will be required to obey $f(\Delta l =
0) = 0$.  We could also require $f(l) - f(l_0)$, but
this will not produce force laws of any better quality and often
the force laws will be more complex in form.

Near the jamming area fraction, our fractional uncertainty is
typically $\delta l/\Delta l_{ij} \approx 100$\$.  In contrast,
ten droplet diameters up the incline ($\phi - \phi_c \approx
0.01$) typical values of $\Delta l_{ij}$ are much larger and the
uncertainty drops to about $\delta l/\Delta l_{ij} \approx 5$\%,
and decreases further still for larger $\phi$.

Next, we wish to know the curvature of the droplet outline.
We measure $r_{i}$ for a droplet by locating each portion of the
perimeter belonging to a water-oil interface, that is, not touching
another droplet.  We fit these portions of the interface to an
arc of constant curvature as shown in Fig.~\ref{fig:rattler}(c,d)
to obtain a local radius of curvature for each portion.  The radii
for the different portions should be the same, but in practice
they vary due to noise.  We arithmetically average these local
radii of curvature to obtain the mean curvature $r_{i}$ for the
droplet. To determine the uncertainty, we create artificial
images of
circular perimeters with known radii and noise commensurate to our
experimental data, and fit these perimeters to find values of $r$.
The distributions of $r$ reproduce the experimentally observed
variance in the individual radii of curvature, 
and let us deduce that the measurement error of $r_i$ is 3.5\% for
the $5\times$ lens and 7\% for the $1.6\times$ lens.  In contrast
to $\Delta l$ which is easier to measure at large $\phi$, $r_c$
is harder to measure at large $\phi$ as the curved portions of
droplets are shorter.  Closer to $\phi_c$, these uncertainties are
smaller.

The last measurable quantity to consider is $\overline{R}$, the
mean 3D curvature of a compressed emulsion droplet in our quasi-2D
system.  This curvature relates to the Laplace pressure and so may
be relevant for the force law, although we will show below that
it is not needed; nonetheless we discuss it for completeness.
For scenarios where droplets are asymmetrically deformed in 3D,
the water-oil interface has two principle radii, the maximum radius
of curvature $R_{i,1}$ and the minimum radius of curvature $R_{i,
2}$. For droplets compressed in this manner, the mean curvature
$1/\overline{R_{i}} = 1/2(1/R_{i, 1} + 1/R_{i,2})$ is constant
anywhere on the surface. 

\begin{figure}
\begin{center}
\includegraphics[width=1.7in]{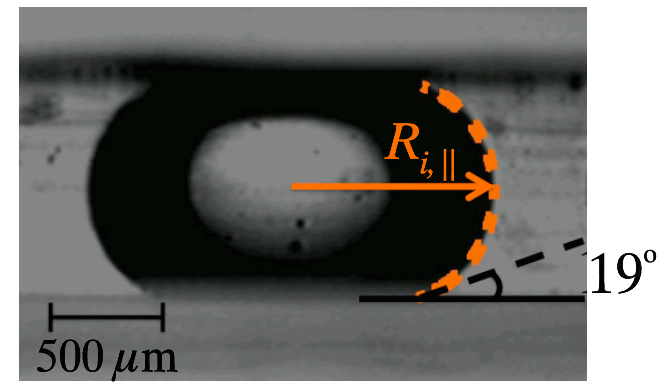}
\end{center}
\caption{(Color online). An experimental image of a
mineral oil droplet squeezed between two glass slides, where the
gap thickness is 1 mm, $R_{i,\|} = 0.88$ mm and $R_{i, \perp} =
0.56$ mm. The orange (light) dashed line is a fit to the
perimeter to obtain $R_{i, \|}$ and $R_{i,
\perp}$.}
\label{fig:finding_R2}
\end{figure}

To measure $R_{i,1}$ and $R_{i,2}$ experimentally we take side view
images of isolated droplets in a sample chamber of gap thickness $h
= 1$ mm (see Fig.~\ref{fig:finding_R2}). The width of the droplet
cross-section is $2 R_{i,1}$, and corresponds to the droplet radius
that would be measured as the 2D radius in the normal top-down view
of our experiments.  The free surface of this compressed droplet is
a surface of mean curvature $\overline{R}$; this is not a circular
arc of constant radius as $R_{1}$ varies with height.  To obtain
$R_{i,1}$ and $R_{i,2}$ of the droplet, we fit the surface using
the the method of Caboussat and Glowinski \cite{Caboussat2008} (an
algorithm to generate the surface of a droplet compressed between
two boundaries). In Fig.~\ref{fig:finding_R2}, we show the fit as
the orange dashed line. Repeating this method for many droplets,
we find $R_{i, \perp}/h = 0.552\pm 0.011$ for droplets in the size
range we use.  For simplicity, we simply use $R_{i,\perp}=0.552h$
for all $r_i$.

\subsection{Mathematical Treatment of an Empirical Force Law}

Our goal is to find an empirical force law
$f(l_{ij},l_0,r_i,r_j)$ relating the contact force between two
droplets $i$ and $j$ to the information about their outlines.
{\it A priori} it is useful to consider what such a force law
should look like.

We first consider two cases where the force law is already
known, the ideal 2D case and the ideal 3D case.  By ideal, we
mean that the contact angle between two droplets is zero, and
where there are no adhesive forces.  Generally these are not
realistic assumptions, due to the interactions between 
the surfactant molecules at
the contacting interface~\cite{Poulin1998, Leal-Calderon2007}. 
For the ideal cases,
the force between two droplets in contact
can be modeled using Princen's 2D model \cite{Princen79, Princen80,
Princen83} or Zhou's 3D model \cite{Zhou2008}. We use lower case to
indicate 2D variables and upper case to indicate 3D variables. In
2D, the contact between two droplets has a contact length $l_{ij}$,
and in 3D, the contact has contact area $A_{ij}$. The force law
for the two models are
\begin{eqnarray}
\mbox{2D Model: } f_{ij} = \gamma_{2D}\frac{l_{ij}}{r_{ij}} \mbox{, where } r_{ij} = \frac{r_{i}+ r_{j}}{r_{i}r_{j}} \\
\mbox{3D Model: } F_{ij} = \gamma_{3D}\frac{A_{ij}}{R_{ij}}  \mbox{, where } R_{ij} = \frac{R_{i}+ R_{j}}{R_{i}R_{j}}
\label{princeneqn}
\end{eqnarray}
In the above equations, $\gamma{}_{2D}$ is a 2D line tension
and $\gamma{}_{3D}$ is a 3D surface tension. For scenarios where
droplets are asymmetrically deformed in 3D, the radius of curvature
$R_{ij}$ in the 3D model must be replaced by the mean curvature
$\overline{R}_{ij}$.

The 2D model would be straightforward to apply as we directly
measure $l_{ij}$, $r_i$, and $r_j$.  To apply the 3D model,
a reasonable assumption is that $A_{ij}$ is related to $l_{ij}$
and perhaps the droplet radii.  The radii $\overline{R}_i$ and
$\overline{R}_j$ are measurable as described in the previous
subsection.  

Rather than choosing between the 2D and 3D models, we test
generalizations of both models and let the data select what
works best.  As described above, one of our variables the force
will depend on is $\Delta l_{ij}$ and we constrain all possible
force laws so that $f(\Delta l_{ij}=0) = 0$.
In general, we consider models of the form
$f_{ij}^{(2D)}(\Delta l_{ij}, 1/r_{ij}; \vec{\alpha})$ for 2D and
$f_{ij}^{(3D)}(\Delta l_{ij}, 1/\overline{R}_{ij}; \vec{\alpha})$
for 3D.  $\vec{\alpha} = \alpha_1, \alpha_2, ...$ are the
fitting parameters associated with a given functional form.
To give an example, we could write $f_{ij}^{2D} = \alpha_1(\Delta
l_{ij}/r_{ij})^{\alpha_{2}}$ with fitting parameters $\alpha_1$ and
$\alpha_2$.  In all, we test a total of 86 various 2D and 3D
force laws of different functional forms that include exponentials,
hertzians, power laws, and polynomials in $l_{ij}$, $1/r_{ij}$,
and $1/\overline{R}_{ij}$, and combinations of these forms.

\subsection{Optimization Problem}

To test the force laws, we establish constraints from the data,
optimize each force law subject to the constraints, and then
quantify how well the optimum force laws describe the data.  
To start with, we consider the constraints
on forces in the $x$ and $y$ directions.

In the $y$-direction the sum of the forces on any given droplet is
equal to the buoyant weight $W_D$.  This is in practice hard to
use directly, as $W_D$ is small compared to the contact forces,
and likely below limits set by noise.  Therefore, rather than
considering individual droplets, we note that droplets located
at a given $y$ must support the observed total buoyant weight
$W_{obs}$ of droplets below them, known simply from measuring the
total area of droplets with centers below $y$.  The way in which
these droplets support this buoyant weight is through contact
forces, and for an assumed force law $f_{ij}(\Delta{}l_{ij},
1/r_{ij}; \vec{\alpha})$ we can determine these contact forces by
substituting our measured values for $\Delta{}l_{ij}$ and $r_{ij}$
(or $\bar{R}_ij$) 
into the function.  If the assumed force model accurately predicts
the forces, then the sum of these contact forces $\sum F_{mod,y}$
at a given $y$ will equal $W_{obs}$.  Here $\sum F_{mod,y}$
are the sum of the $y$-component of only those forces pointed in
the downward directions.  The reason we only consider the downward
facing forces is because the collective buoyant weight is pushing
upward, and to satisfy Newton's 3rd law, the balancing forces must
be facing downward.  We convert $W_{obs}$ and $F_{mod,y}$ into
2D pressures (force per unit length) by writing $\lambda_{obs} =
W_{obs}/w$, $\lambda_{mod} = \sum F_{mod,y}/w$, using the width
of the chamber $w$.  $\lambda$ is in essence the 2D hydrostatic
pressure at height $y$.  Because there is no static friction at
the sidewalls, there is no Janssen effect~\cite{Janssen1895}.

We define a goodness of comparison in the $y$-direction as
\begin{equation}
\chi_y^2 = \displaystyle\sum_y \left[\left(\lambda(y)_{obs} -
\lambda(y)_{mod}\right)/\langle{}\lambda(y)_{obs}\rangle{}\right]^2,
\end{equation}
where smaller values of $\chi_y^2$ indicate a better match
between the assumed force law and the actual forces. In
the equation, $y$ indexes various distances up the incline
where $\lambda(y)_{mod}$ and $\lambda(y)_{obs}$ are sampled,
and the angle brackets are an average over $y$. We normalize
by $\langle{}\lambda(y)_{obs}\rangle$  to make $\chi_y^2$
dimensionless, and since $\langle{}\lambda(y)_{obs}\rangle{}$
is independent of the assumed force law, it does not change the
results. We sample $\lambda$ at intervals of 5$r_{0}$ up the
incline. At each $y$ sampled, $\lambda_{mod}$ is calculated using
the contact lengths and droplet radii for all droplets found
between a distance $y - 5r_{0}$ and $y+5r_{0}$ up the incline,
and $\lambda_{obs}$ is calculated using the position and radii of
all droplets below a distance $y$ up the incline.

We next consider the forces in
the $x$-direction.  In contrast to the $y$-direction there are
no external forces, so the sum of the forces on each droplet in
the $x$-direction is zero. From this we construct the goodness
of comparison
\begin{equation}
\chi_x^2 = \displaystyle\sum_i \left[\left(\displaystyle\sum_j f_{x, ij}  \right)/\langle{}|\vec{f}_i|\rangle\right]^2,
\end{equation}
where the $F_{x, ij}$ is the $x$ component of the force
at a contact between droplets $i$ and $j$ and
$\langle{}|\vec{f}_i|\rangle$ is the average net contact force
exerted on droplet $i$. In the equation, $f_{x, ij}$ are the forces
predicted by the assumed force law. Due to measurement error, the
forces will not sum to zero, and the deviation from zero grows
with $\langle{}|\vec{f}_i|\rangle$. We assume that the deviation
will grow linearly with $\langle{}|\vec{f}_i|\rangle$ and to
fairly weight the contributions of each droplet to $\chi_x^2$,
we normalize the sum of the forces by $\langle{}|\vec{f}_i|\rangle$.

Finally, we define a net goodness of comparison $\chi^2 =
\chi_x^2\chi_y^2$ which indicates how well an assumed force
law models the forces in both the $x$ and $y$ directions. Since
we know the buoyant weight of our droplets in units of $\mu$N,
this allows us to find a force law in units of $\mu$N. Later,
we compare $\chi^2$ between the different force laws to determine
the best overall force law.

\begin{table}[t]
\begin{center}
\begin{tabular}{ccccc}
\hline
$R_{0}$ [$\mu$m] & $r_{0}$ [$\mu$m] & $h$ [$\mu$m]  & poly (\%) & $l_{0}$ [$\mu$m]  \\
\hline
164 &  183 &186 & 5.5 & 50.8 \\
143 & 156 & 180  & 1.6  &  48.8 \\
105 &  128 &  96 & 1.9 & 36.6 \\
84.1 &   89 & 96  & 2.5 & 30.3 \\
\hline
\end{tabular}
\end{center}
\caption{Parameters characterizing the droplets in our 4 different
monodisperse samples. $R_{0}$ is the 3D radius of the droplets,
$r_{0}$ is the 2D radius, $h$ is the gap thickness of the chamber,
poly is the polydispersity of the sample, and $l_{0}$ is the length
of contact for two droplets just in contact. Our uncertainties
for the various measures are $\pm$ 0.2 $\mu$m in $R_{0}$, $\pm$
2 $\mu$m in $r_{0}$, and $\pm$ 4 $\mu$m in $h$. The variability
in $l_{0}$ grows with droplet size and the measurement uncertainty
can be expressed as $\pm$ 0.04$r_{0}$ in $l_{0}$.}
\label{table:Mono}
\end{table}

\subsection{Empirical Force Law: Monodisperse and Bidisperse}

We now apply our method to find an empirical force law. We start
by determining the force law for same size droplets in contact
using data taken on four different monodisperse samples. The
samples are prepared by placing droplets with 3D radius $R_{0}$
into a sample chamber with gap thickness either $100 \pm 4$
$\mu$m or $180 \pm 4$ $\mu$m, and once in the chamber, the
droplets have a 2D radius of $r_{0}$.  The error assigned in the gap
thickness represents the unavoidable variations in the gap thickness
measured at different points along the sample chamber.  We note
that the larger the sample chamber, the larger the variability of
gap thickness we observe.  As the chambers are inexpensive, sample
chambers with more variability than $\pm 4$~$\mu$m are discarded.
After the sample chambers are filled, they are sealed to prevent
evaporation, and then placed on a microscope inclined at
28$^{\circ}$.  Droplets rise to the top and come to rest in
mechanical equilibrium, at which point we acquire images of the
sample. Various parameters characterizing each
monodisperse sample are shown
in Table~\ref{table:Mono}.

To determine the best force law we pick each possible functional
form, optimize the parameters $\vec{\alpha}$ for it, and then
compare $\chi^2$ for the different functions.  Several functional
forms all have small $\chi^2$ values, and of these we choose
one that is simple and plausible.  For simplicity, 
measuring the 2D $r_{ij}$ is simpler than measuring the
3D $\overline{R}_{ij}$.  For plausibility, functions that treat
$\Delta{}l_{ij}$ and $r_{ij}$ as $\Delta{}l_{ij}/r_{ij}$ most
closely resemble Eqn.~\ref{princeneqn}.

We judge the most reasonable
function with low $\chi^2$ to be 
$F_{ij} = \alpha_1\Delta{}l_{ij}/r_{ij} +
\alpha_2(\Delta{}l_{ij}/r_{ij})^2$, quite similar to
Eqn.~\ref{princeneqn}.  We
also choose this function because we find that we can easily rescale
this function using the 3D droplet radius $R_0$ so that we have a
universal force law for all four data sets (the four different
droplet sizes).  The rescaled force law is 
\begin{equation} 
F_{ij} = \alpha_1R_{0}\Delta{}l_{ij}/r_{ij} +
\alpha_2(R_{0}\Delta{}l_{ij}/r_{ij})^2.  
\label{eq:SameSizeForceLaw}
\end{equation}
where $\alpha_{1} = 4.25$~$\mu$N/mm and $\alpha_{2} =
4.12$~$\mu$N/mm$^2$.  

\begin{figure}
\includegraphics[width=3.3in]{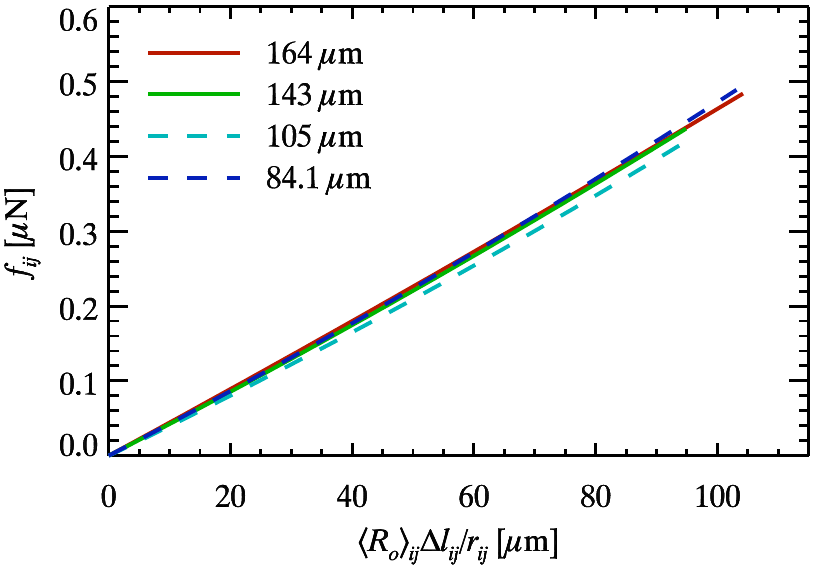}
\caption{(Color online). A plot of the universal empirical force law for each data set. The solid lines are samples with a gap thickness of approximately $180$ $\mu$m and the dashed lines are samples with a gap thickness of approximately $110$ $\mu$m. The legend indicates $R_{0}$.}
\label{fig:monofit}
\end{figure}

The rescaled force law is shown for each
data set in Fig.~\ref{fig:monofit}. We see that all the force
laws are nearly linear; the quadratic correction is about 10\%
for the largest forces.  The rescaled force law shows that all
the data collapse very well and only slightly deviate between each
other at larger $\Delta{}l_{ij}/r_{ij}$.  For these larger values in
$\Delta{}l_{ij}/r_{ij}$, the area fraction is close to 0.96 which is
the upper limit where we can still confidently measure $r_{ij}$;
for those close-packed droplets, only a few pixels occupy the
water-oil interface and $r_{ij}$ is hard to determine.  We do not
claim that Eqn.~\ref{eq:SameSizeForceLaw} is the correct force law;
in particular, while $\alpha_1$ has units of surface tension and is
plausible for an oil-water surface tension, the physical meaning
of $\alpha_2$ is unclear.  Rather, Eqn.~\ref{eq:SameSizeForceLaw}
accurately provides the forces between our droplets, within the
measurement limitations set by our data. Also, there may be other sources of error, for instance, Lacasse \textit{et al.}~\cite{Lacasse1996} has numerically shown that the force law has a slight sensitivity to the number of neighbors and the relative positioning of the neighboring droplets. 
To examine if there are other potential sources of error, using
Eqn.~\ref{eq:SameSizeForceLaw} we compared the deviations in the
computed net force on each droplet to the deviations we expect
given our measurement errors, and find the two agree well.  Thus,
within the limitations of our measurement errors, we have
resolved the forces as best as possible, confirming
Eqn.~\ref{eq:SameSizeForceLaw} is adequate.




\begin{table}
\begin{center}
\begin{tabular}{ccccccc}
\hline
$\sigma$ & $R_{0}^{(big)}$ & $R_{0}^{(small)}$ & $h$ & $l_{0}^{(ss)}$ & $l_{0}^{(sb)}$ & $l_{0}^{(bb)}$ \\
 & [$\mu$m] & [$\mu$m] & [$\mu$m] & [$\mu$m] & [$\mu$m] & [$\mu$m] \\
\hline
1.25 & 102 & 86.5 & 104 & 52.0 & 56.0 & 63.0  \\
1.42 & 91.8 & 80.0 & 106 & 46.8 & 50.9 & 58.5 \\
1.52 & 108 & 79.2 & 104 & 45.0 & 51.0 &  58.7 \\
\hline
\end{tabular}
\end{center}
\caption{The parameters characterizing the
droplets in our 3 bidisperse samples. Other parameters related to these
droplets are shown in Table~\ref{table:Binary1}. The first column
$\sigma = r_{0}^{(big)}/r_{0}^{(small)}$ is the size ratio. The two
radii shown are the 3D droplet radii of the small and big droplets
before placing them in the chamber. There are also three contact
types:  small-small ($ss$), small-big ($sb$),
and big-big ($bb$), and therefore there are three $l_{0}$
values. Our uncertainties for the various measures are $\pm$ 0.2
$\mu$m in $R_{0}$ and $\pm$ 4 $\mu$m in $h$. The variability in
$l_{0}$ grows with droplet size and the measurement uncertainty
can be expressed as $\pm$ 0.04$r_{0}$.}
\label{table:Binary}
\end{table}

To test how well one can determine a force law given finite data
and measure error, we additionally simulated inclined mechanically
stable droplet packings of 1000 droplets with a known force law,
then added noise to the data consistent with experimental noise.
Applying our empirical method to the simulated data, we recover
the known force law with 2\% errors in the coefficients
(noise equivalent to the experiments using the $5\times$
microscope objective) or 5\% errors in the coefficients (noise
equivalent to the experiments with the $1.6\times$ microscope
objective).  
This suggests it is possible that the $\sim 4$\%
variations between the force laws for different sized droplets
seen in Fig.~\ref{fig:monofit} are simply due to noise, and they
may well have exactly the same force law.

So far we have focused on force laws in monodisperse samples, but
we also need to measure forces between different-sized droplets
in bidisperse samples.  To obtain a force law between droplets of
different sizes, we apply our method to find an empirical force
law using data taken on three different bidisperse samples. The
bidisperse samples are prepared in the same manner as the
monodisperse case.  Table~\ref{table:Binary} summarizes
the various parameters of our bidisperse systems; see also
Table~\ref{table:Binary1}.

For the case of a bidisperse sample with small and big droplets, there
are 3 possible contact types to consider:  small-small, small-big,
and big-big.  Our previous results give us small-small and big-big
forces.  We assume the unknown small-big
force law obeys the same functional
form as the monodisperse case (Eqn.~\ref{eq:SameSizeForceLaw}),
where $\alpha_{1}$ and $\alpha_{2}$ need to be determined.
Recall that Eqn.~\ref{eq:SameSizeForceLaw} contains a term $R_{0}$
that rescales the force law and makes it universal. For the
small-big contacts there are two different $R_0$ values, one for
each droplet size.  To account for these two radii we 
substitute $R_{0}$ with the arithmetic mean of the two
radii $\langle{}R_{0}\rangle{}$ giving as our bidisperse empirical force
law $F_{ij} = \alpha_1\langle{}R_{0}\rangle{}\Delta{}l_{ij}/r_{ij}
+ \alpha_2(\langle{}R_{0}\rangle{}\Delta{}l_{ij}/r_{ij})^2$, where
$\alpha_{1}$ and $\alpha_{2}$ are unknown. To obtain $\alpha_{1}$
and $\alpha_{2}$ for our bidisperse samples we minimize $\chi_x^2$,
and find that $\alpha_{1}$ and $\alpha_2$ are very close to
that found for the monodisperse case and within the 4\% variation we expect from due to finite sampling and measurement error. Therefore, we have shown that to within 5\% error we have a found a universal force law that works for any droplet size and is close to Princen's 2D model \cite{Princen79, Princen80, Princen83} with a small second order correction.


\subsection{Final comments on force law}

The uncertainties in determining forces are related to the
magnification.  The higher the magnification, the better we
can measure the contact length $l$ and the mean curvature $r$.
Fortunately, given that we 
study static samples, this means we can take overlapping images
at high magnification to reduce our uncertainties, as described
in Sec.~\ref{method}.  In an
experiment with moving droplets, overlapping images of different
fields of view would be difficult or impossible.  This situation
would require limiting the field of view to fewer droplets, if the
same resolution of forces was desired.  For any magnification,
uniformity of lighting is essential so that the appearance of
droplets is uniformly related to their true shape and size.
As discussed in the previous subsection, our imperfect knowledge
of the force law gives us a systematic uncertainty no worse
than 5\%.  To determine the random uncertainty for particular
forces, we take measured $r_{ij}$ and $\Delta l_{ij}$ values,
add noise commensurate to our known uncertainty (discussed in
Sec.~\ref{measurablevariables}), and recalculate the force to
see the variation.  The bidisperse data of Sec.~\ref{jamming} were
taken with a $1.6\times$ lens and have a random uncertainty of 16\%.
The monodisperse data used in the calibration procedure were taken
with a $5\times$ lens and have a random uncertainty of 8\%.

An additional experimental complication is that droplets at
rest occasionally feel a static force from the top and bottom
plates.  This is likely due to contact line pinning on
impurities or microscopic scratches on the glass.  To minimize
this, we pre-clean each slide with methanol which we gently blow
off the slide.  Harsher cleaning methods do not significantly
reduce the droplet pinning.  The magnitude of these forces can
be estimated by examining a dilute concentration of droplets
in a horizontal slide, and then slowly tilting the slide to see
when the droplets begin to move due to gravity.  For the samples
discussed in this work, they begin to move at tilt angles of
about 4.5$^{\circ{}}$ or sooner.  We discard any sample chambers
with pinning stronger than this.  Given that our experiments are
conducted at a tilt angle of 28$^\circ$, the buoyant weight of a
droplet is $\sin(28^\circ)/\sin(4.5^\circ) = 6$ times larger than
any pinning force. Taking the analysis a little further, since the friction force on any droplet can range from zero to the maximum, a more appropriate estimate for the buoyant weight of a
droplet is $\sin(28^\circ)/(\sin(4.5^\circ)/2) = 12$ times the average pinning force. These estimates show that the inter-droplet forces seen in the jammed
emulsions ($\phi > \phi_c$) are on the order of a hundred times the pinning forces.
We believe that the situation in our calibration experiments are even more favorable.
After compaction, the pinning forces should be in random directions,
as an analogous granular experiment observed that particles move
in random directions during compaction \cite{Pouliquen2003}. Therefore, a vector average gives a pinning force on each droplet very close to zero. Since our empirical method relies on the average vector forces on a droplet, the pinning forces can safely be neglected in the results
of Sec.~\ref{jamming}.  
Note that if our experiment was scaled down in size (smaller
droplets, thinner plate gap) the pinning forces become more
significant compared to the droplet weight and can dominate the results.


\footnotesize{
\bibliography{all_ref2} 
\bibliographystyle{rsc} 
}

\end{document}